\documentclass[]{revtex4-2}
\usepackage[utf8]{inputenc}
\usepackage{graphicx} 
\usepackage{xcolor}
\usepackage{physics}
\usepackage{dutchcal}
\usepackage{booktabs} 
\graphicspath{{ARXIV/}}

\usepackage{soul} 
\DeclareMathAlphabet\mathsf{OT1}{lcmss}{m}{n}

\begin{document}

\title{Nonlinear dynamics of air invasion in one-dimensional compliant fluid networks}

\author{Ludovic Jami}
\email{jami.ludovic@gmail.com}
\affiliation{Université Côte d’Azur,  CNRS UMR 7010, Institut de Physique de Nice, 17 rue Julien Lauprêtre, 06200 Nice, France}
\author{François-Xavier Gauci}
\affiliation{Université Côte d’Azur,  CNRS UMR 7010, Institut de Physique de Nice, 17 rue Julien Lauprêtre, 06200 Nice, France}
\author{Céline Cohen}
\affiliation{Université Côte d’Azur,  CNRS UMR 7010, Institut de Physique de Nice, 17 rue Julien Lauprêtre, 06200 Nice, France}
\author{Xavier Noblin}
\affiliation{Université Côte d’Azur,  CNRS UMR 7010, Institut de Physique de Nice, 17 rue Julien Lauprêtre, 06200 Nice, France}
\author{Ludovic Keiser}
\email{ludovic.keiser@univ-cotedazur.fr}
\affiliation{Université Côte d’Azur,  CNRS UMR 7010, Institut de Physique de Nice, 17 rue Julien Lauprêtre, 06200 Nice, France}

\begin{abstract}
Vascular networks exhibit a remarkable diversity of architectures and transport mechanisms across biological systems. Inspired by embolism propagation in plant xylem, where air invades water-filled conduits under negative pressure, we study air penetration in compliant one-dimensional hydrodynamic networks experiencing mass loss by pervaporation. Using a theoretical framework grounded in biomimetic models, we show that embolism dynamics are shaped by the interplay between network compliance and viscous dissipation. In particular, the competition between two timescales (the pressure diffusion time, $\tau_\mathrm{diff}$, and the pervaporation time, $\tau_\mathrm{pv}$) governs the emergence of complex, history-dependent behaviors. When $\tau_\mathrm{diff} \sim \tau_\mathrm{pv}$, we uncover a nonlinear feedback between the internal pressure field and the embolism front, leading to transient depressurization and delayed interface motion. These results offer a minimal framework for understanding embolism dynamics in slow-relaxing vascular systems and provide design principles for soft microfluidic circuits with tunable, nonlinear response.
\end{abstract}

\maketitle

\section{Introduction}

Vascular networks are essential architectures in living organisms, spanning a wide range of geometries and transport strategies \cite{wootton1999fluid,marbach_vein_2023,le_verge-serandour_physarum_2024, oyarte_galvez_travelling-wave_2025, ogawa2025architecture, jensen_sap_2016}.  From distributing nutrients and signaling molecules \cite{alim_mechanism_2017} to maintaining homeostasis, their functions rely on designs that often reconcile transport efficiency \cite{liese_balancing_2021} with robustness \cite{pereira_angiosperms_2023}. These networks frequently exhibit a coupling between flow and elasticity, termed hydraulic compliance, where the fluid storage capacity depends on local pressure \cite{heil_fluid-structure_2011}. 

This rich interplay has inspired the development of artificial fluidic networks, especially in the context of flexible microfluidics \cite{fallahi2019flexible}. These systems have revealed a diversity of nonlinear dynamics \cite{battat_nonlinear_2022,xia_nonlinear_2021}, triggered by inertial effects \cite{case_spontaneous_2020}, nonreciprocal elements \cite{brandenbourger_tunable_2020,park_fluid-structure_2021}, or active boundaries such as cilia \cite{jambonpuillet2025densearrayelastichairs}. Such mechanisms can lead to complex and sometimes counterintuitive behaviors (e.g., the emergence of the Braess paradox \cite{case_braesss_2019} or fluidic memory elements \cite{martinez-calvo_fluidic_2024}) highlighting the potential of nonlinear fluid-structure interactions. In confined geometries, both elasto-capillary \cite{bradley_wettability-independent_2019, bradley_bendocapillary_2023, ushay2023interfacial} and elastohydrodynamic couplings \cite{pihler-puzovic_suppression_2012, juel_instabilities_2018, louf_bending_2020, cappello_fiber_2022,guyard2022elastohydrodynamic,paludan_elastohydrodynamic_2024,garg_passive_2024} have been harnessed to design responsive and/or adaptable transport systems.

Among natural vascular systems, the xylem in plants is particularly intriguing \cite{tyree_xylem_2013, stroock_physicochemical_2014, katifori_transport_2018}. It sustains upward water transport from roots to leaves, driven by negative pressure gradients established via leaf transpiration. Some species tolerate tensions down to $-10$ MPa, yet drought-induced embolism (the entry and spread of gas into xylem conduits) can severely disrupt water transport, leading to plant decline or death \cite{choat2012global, choat2018triggers,torres2024plant}. In leaves, embolism propagation has been directly visualized using high-resolution imaging techniques like the “optical vulnerability method” \cite{brodribb_revealing_2016, brodribb_optical_2017}, revealing an intermittent pattern, possibly governed by bordered pits acting as capillary barriers. Furthermore, structural failure, such as wall buckling or collapse, under extreme negative pressures has been documented \cite{cochard_xylem_2004, zhang_reversible_2016, chin_tracheid_2022}, pointing to a role of fluid-structure interactions in embolism dynamics. 
However, despite progress in imaging, the multiscale and rapid nature of embolism propagation makes it difficult to access the full pressure dynamics experimentally.

To bridge this gap, biomimetic approaches have been developed to replicate xylem behavior in controllable artificial settings. Early models used hydrogels or silicone networks to study transpiration \cite{wheeler_transpiration_2008, noblin_optimal_2008}, cavitation \cite{duan_evaporation-induced_2012,vincent_birth_2012, vincent_drying_2014, bruning_turning_2019}, and air-seeding processes \cite{dollet_drying_2019, dollet_drying_2021,vincent_tunable_2024}. More recently, researchers have mimicked bordered pits with microfluidic constrictions, imposing Laplace pressure thresholds that temporarily block air invasion until pressure relaxation permits sudden jumps \cite{keiser_intermittent_2022}. The elasticity of the channels plays a key role, modulating how pressure evolve dynamically with volume loss by pervaporation (mimicking evapotranspiration).

\begin{figure*}[t!]
    \centering
    \includegraphics[width=0.9\textwidth]{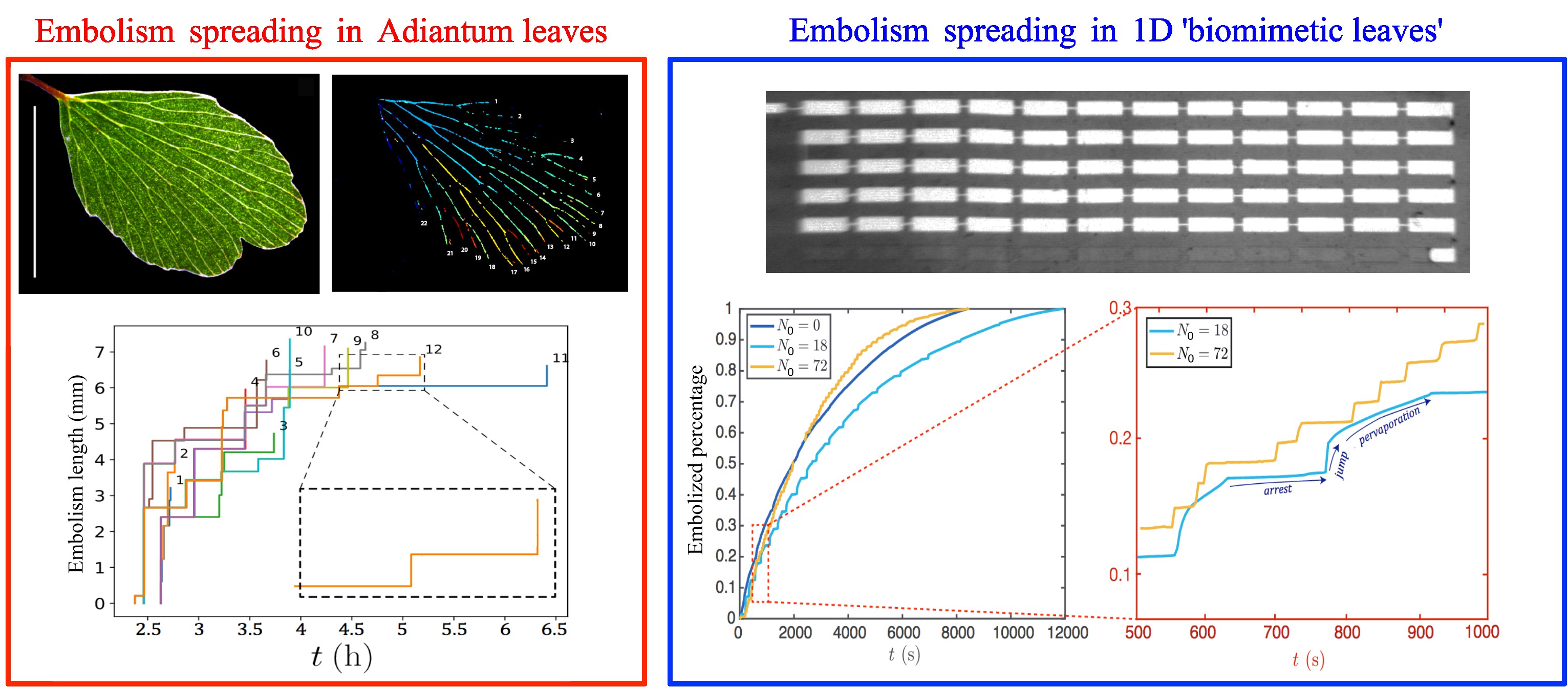}
    \caption{Left: Picture of the embolism propagation in an \textit{Adiantum} leaf (scale bar = 5 mm) and embolism growth as a function of time for different veins. Courtesy of the authors of Ref. \cite{keiser_embolism_2024}, where the analysis was carried out based on the experiments from Brodribb \textit{et al.} \cite{brodribb_revealing_2016}. For each vein, the dynamics is highly intermittent, with sudden sub-minute propagation events followed by hour-long periods of rest. Right : Embolism dynamics in an one-dimensional pervaporating biomimetic network made with PDMS (seen from the top), where analogous intermittent propagations were evidenced. Courtesy of A. Pellegrin for the picture (whose vertical dimension represents 4 mm), and of the authors of Ref. \cite{keiser_embolism_2024} for the data. The white vessels are embolized (full of air) while the grey ones are still full of water. Note that the curves representing the embolized percentage of the networks exhibit a concave shape. This is a signature of a coherent network (regardless of the number of channels $N_0$) where pressure variations are rapidly transmitted throughout the structure. The purpose of our current study is to explore configurations where pressure diffusion is slower and where networks are less coherent.}
    \label{fig:Adiantum}
\end{figure*}

Recent work has shown that these principles can be extended to channel arrays inspired by \textit{Adiantum} leaves, reproducing intermittent embolism dynamics consistent with real leaves (Fig.~\ref{fig:Adiantum}, \cite{keiser_embolism_2024}). Yet, existing studies consider uniform liquid pressure and focus on pervaporation or capillarity, neglecting viscous dissipation. This assumption breaks down in networks combining wide channels (producing large evaporation fluxes) and narrow constrictions (with high hydraulic resistance), where hydrodynamic pressure variations can rival or exceed capillary thresholds.

In this work, we develop a model of embolism propagation in compliant, xylem-inspired, one-dimensional fluidic networks with significant hydrodynamic resistance. While the system is deliberately idealized and not intended to quantitatively replicate plant xylem, it contains a minimal set of physical ingredients (compliance, pervaporation, and capillary thresholds) relevant to embolism dynamics in synthetic and biological conduits alike. 

By capturing the coupling between pervaporation-induced flows, Laplace thresholds, and viscous pressure drops, we explore how key design parameters govern the dynamics. In particular, we demonstrate that when the pressure diffusion timescale is comparable to or larger than the pervaporation timescale, a nonlinear feedback emerges between the embolism front position and the internal pressure field. This coupling leads to complex, history-dependent dynamics, where embolism progression is no longer solely dictated by local capillary pressure thresholds but also by the evolving pressure distribution across the network. This framework offers interesting perspectives on embolism spreading in plants, where pressure heterogeneity may naturally arise from long transport paths and anatomical complexity. It also informs the design of soft, nonlinear fluidic circuits with emergent behaviors.

The paper is structured as follows. In Sec. \ref{sec:description}, we describe the physical model and the architecture of the networks. In Sec. \ref{sec:model}, we present the numerical resolution in both discrete (Subsec. \ref{subsec:discrete}) and continuous (Subsec. \ref{subsec:continuous}) settings, and analyze how network topology influences pressure drops (Subsec. \ref{subsec:design}). In Sec. \ref{sec:discussion}, we identify asymptotic regimes where pervaporation dynamics are slow (Subsec. \ref{subsec:larger}) or fast (Subsec. \ref{subsec:smaller}) compared to pressure diffusion. We finally discuss the broader implications for nonlinear transport in engineered fluidic networks (Subsec. \ref{subsec:fluidic_networks}) and plant physiology (Subsec. \ref{subsec:plants}).

\section{Description of the system: Pervaporation, elasto-capillary and hydrodynamic coupling of the channels.}
\label{sec:description}

\begin{figure*}[t]
    \centering
    \includegraphics[width=0.9\textwidth]{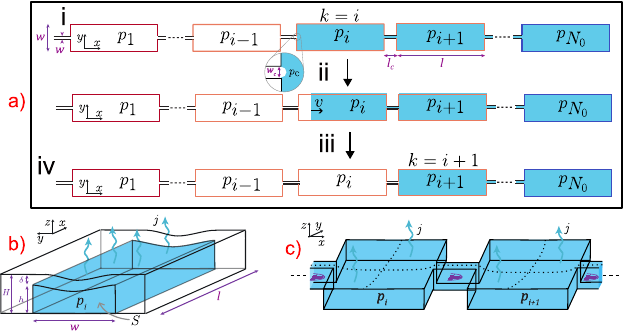}
    \caption{\textbf{Illustration of a series of compliant pervaporating channels connected by constrictions.} (a) Illustration of the embolism propagation in the channel series. The width and length of the channels $w$ and $l$ and of the constriction are indicated, together with the pervaporation rate $j$ whose main contribution comes from the upper thin wall. In step $\mathsf{i}$, the embolism front is halted at the constriction between channels $i-1$ and $i$. To cross the constriction (step $\mathsf{ii)}$, pressure $p_{k+1}$ must drop below the Laplace pressure imposed by the constriction (eq. \ref{eq:Laplace_p_c}). In step $\mathsf{iii}$, the embolism has fully invaded channel $i$, and the interface reaches the next constriction. Step $\mathsf{iv}$ is equivalent to step $\mathsf{i}$ but with the embolism front, indexed by $k$, now located between channels $i$ and $i+1$. (b) Scheme of the deformation of the channel when loosing water volume by evaporation. In this scheme, the channel dimensions $h$, $w$, $l$ and $\delta$ are defined, as well as the section of the deformed channel $S = S_0 + \Delta S$ with $S_0=hw$ the section at rest. (c) Pressure difference between two successive channels $i$ and $i+1$ generates a flow through the connecting constriction.}
    \label{fig:serie}
\end{figure*}
We investigate series of $N_0$ microchannels fabricated in PDMS, each of length $l$, width $w$ and height $h$, separated by constrictions of length $l_c$, width $w_c$ and height $h_c$ (Fig.~\ref{fig:serie}). In the numerical simulations, we restrict ourselves to constrictions having the same height as the channels ($h=h_c$), but in analytical expressions we retain distinct notations to clearly separate their geometric contributions. The total thickness of the PDMS leaf $H$ is uniform, such that the thickness separating the ceiling of the microchannels and constrictions from the external air is $\delta = H-h$. The configuration is a dead-end series: the last ($N_0^{\text{th}}$) channel is closed by a wall, and the first constriction being connected to the ambient air by an opening. The thickness $\delta$ is thin enough to enable a relatively fast pervaporation flux. For water-permeable materials like PDMS, the evaporation flux per unit length is given, following Dollet \textit{et al.}, by \cite{dollet_drying_2021}:
\begin{equation}
    j = j_0\qty{\frac{w}{\delta} + \frac{2}{\pi}\qty[\ln(\frac{(H+\delta)h}{\delta^2})+ \frac{H}{\delta}\ln\qty(\frac{H+\delta}{h})]},
    \label{eq:pervap}
\end{equation}
with $j_0 = \frac{M}{\rho}D_\mathrm{p}c_\mathrm{p}^\mathrm{sat}(1-RH)$, where $h$ is the channel height, $H = \delta + h$, $M$ and $\rho$ are, respectively, the molar mass of water and the density of liquid water, $D_\mathrm{p}$ and $c_\mathrm{p}^\mathrm{sat}$ are the diffusivity and saturation concentration of water vapor in the pervaporating material and $RH$ is the relative humidity of the air surrounding the channel network. The second term in the square brackets of equation (\ref{eq:pervap}) is a geometrical logarithmic correction term to account for cases when the height of the channel is comparable to the channel width and similar or greater than the thickness of its upper wall (that is when $h \sim w$ and $h \gtrsim \delta$).
Pervaporation causes a gradual loss of liquid volume, leading to several possible consequences. 
\begin{itemize}
    \item[-] In the absence of a gas phase, volume loss translates into a pressure drop within the channel.  For soft materials like PDMS, this causes an elastic deformation of the upper wall. To describe the relationship between the water volume lost by the channel $\Delta V_i$ and its pressure decrease $p_i$, we define the compliance $C$ as: $\Delta V_i = Cp_i$. We here use the notation $p_i$ for the difference between the pressure in the channel $i$ with the atmospheric pressure in the air surrounding the PDMS leaf. In the case of rectangular channels with thin upper wall ($\delta \ll w$), thin plate elastic theory enables to express the compliance as a function of the geometric dimensions of the channel and the mechanical parameters of the elastic material \cite{keiser_intermittent_2022, keiser_embolism_2024}: 
\begin{equation}
C = \frac{lw^5}{\delta^3}\frac{1-\nu^2}{60E},
\label{eq:compliance}
\end{equation}
with $E$ the Young modulus of the channel material and $\nu$ its Poisson's ratio. In practice, we take $\nu = 0.5$, a standard value for nearly incompressible PDMS elastomers.
 \item[-] When a gas bubble is present, pressure differences between adjacent channels drive water flow through the constrictions. The flow rate between channel $i$ and $i+1$ is governed by the hydrodynamic resistance $R$ of the constriction $q_i=\frac{p_{i+1}-p_i}{R}$. Assuming $w_\mathrm{c}<h_\mathrm{c} = h$, the hydrodynamic resistance is expressed as \cite{bruus_theoretical_2008}:
\begin{equation}
    R = \frac{12\eta l_\mathrm{c}}{h_\mathrm{c} w_\mathrm{c}^3}\frac{1}{1-0.63\frac{w_\mathrm{c}}{h_\mathrm{c}}},
    \label{eq:hydro_R}
\end{equation}
with $\eta$ the dynamic viscosity of the water. 
\end{itemize} 

When embolism propagates across the channel series, the interface can only cross a constriction if the pressure in the downstream channel drops below a Laplace pressure threshold $p_c$. For rectangular constrictions, a completely wetting liquid and assuming $w_\mathrm{c} < h_\mathrm{c}$, the threshold Laplace pressure is given by \cite{wong1992three}:
\begin{equation}
    p_\mathrm{c} = -\gamma \kappa_\mathrm{c},
        \label{eq:Laplace_p_c}
    \end{equation}
with the curvature: 
\begin{equation}
\kappa_\mathrm{c} = \frac{1}{w_\mathrm{c}}+ \frac{1}{h_\mathrm{c}}+\sqrt{\left(\frac{1}{w_\mathrm{c}}-\frac{1}{h_\mathrm{c}}\right)^2+\pi\left(\frac{1}{h_\mathrm{c}w_\mathrm{c}}\right)},
    \label{eq:kappa_c}
\end{equation}
and $\gamma$ the surface tension of water.

\begin{table*}[t]
\centering
\begin{tabular}{@{}lllcl@{}}
\toprule
\textbf{Category} & \textbf{Parameter} & \textbf{Symbol} & \textbf{Typical value (unit)} & \textbf{Fixed / Explored} \\
\midrule
\textbf{Geometric (channels)} \\
 & \textbf{Number of channels} & $N_0$ & $10$ or $64$ & Explored \\
 & \textbf{Channel width} & $w$ & $50$–$600\,\mu\mathrm{m}$ & Explored \\
 & \textbf{Channel length} & $l$ & $0.5$ or $5\,\mathrm{cm}$ & Explored \\
 & \textbf{Channel height} & $h$ & $50\,\mu\mathrm{m}$ & Fixed \\
 & \textbf{PDMS ceiling thickness} & $\delta$ & $H - h$ & Fixed \\
\addlinespace[0.5em]
\textbf{Geometric (constrictions)} \\
 & \textbf{Constriction width} & $w_c$ & $1$–$100\,\mu\mathrm{m}$ & Explored \\
 & \textbf{Constriction length} & $l_c$ & $0.3$–$1000\,\mathrm{mm}$ & Explored \\
 & \textbf{Constriction height} & $h_c$ & $50\,\mu\mathrm{m}$ & Fixed \\
\addlinespace[0.5em]
\textbf{Material properties} \\
 & \textbf{Young’s modulus of PDMS} & $E$ & $1\,\mathrm{MPa}$ & Fixed \\
 & \textbf{Poisson’s ratio of PDMS} & $\nu$ & $0.5$ & Fixed \\
\addlinespace[0.5em]
\textbf{Fluid properties} \\
 & \textbf{Dynamic viscosity of water} & $\eta$ & $1\times10^{-3}\,\mathrm{Pa\cdot s}$ & Fixed \\
 & \textbf{Surface tension of water} & $\gamma$ & $72\times10^{-3}\,\mathrm{N/m}$ & Fixed \\
 & \textbf{Density of water} & $\rho_\mathrm{w}$ & $1000\,\mathrm{kg/m^3}$ & Fixed \\
 & \textbf{Molar mass of water} & $M$ & $18\,\mathrm{g/mol}$ & Fixed \\
\addlinespace[0.5em]
\textbf{Pervaporation parameters} \\
 & \textbf{Diffusivity of water in PDMS} & $D_\mathrm{p}$ & $1\times10^{-9}\,\mathrm{m^2/s}$ & Fixed \\
 & \textbf{Saturation conc. in PDMS} & $c_\mathrm{p}^\mathrm{sat}$ & $40\,\mathrm{mol/m^3}$ & Fixed \\
 & \textbf{Relative humidity} & $RH$ & $0\,\%$ & Fixed \\

\bottomrule
\end{tabular}
 \caption{Physical and geometrical parameters used in the model, grouped by category.}

\label{tab:parameters}
\end{table*}
Eqs. (\ref{eq:pervap})-(\ref{eq:Laplace_p_c}) provide a consistent description of embolism dynamics under a set of reasonable assumptions. Pressure is assumed homogeneous within each channel owing to the high contrast in viscous dissipation between channels and constrictions (Eq. \ref{eq:hydro_R}). Contact line pinning and additional contact line dissipation are neglected. An inertialess Stokes flow regime is considered. Finally, the channel deformation is treated within linear thin plate theory (Eq. \ref{eq:compliance}).

The full range of parameter values used in this study is summarized in Table \ref{tab:parameters}. In the following sections, embolism dynamics will be described in general terms, without making further assumptions on the functional forms of $R$, $C$ $j$ and $p_\mathrm{c}$.

\section{Embolism propagation and pressure dynamics in one-dimensional networks}
\label{sec:model}

\subsection{Embolism dynamics in a discrete network of channels and constrictions in series.}
\label{subsec:discrete}

\begin{figure*}[t]
    \centering
\includegraphics[width=\textwidth]{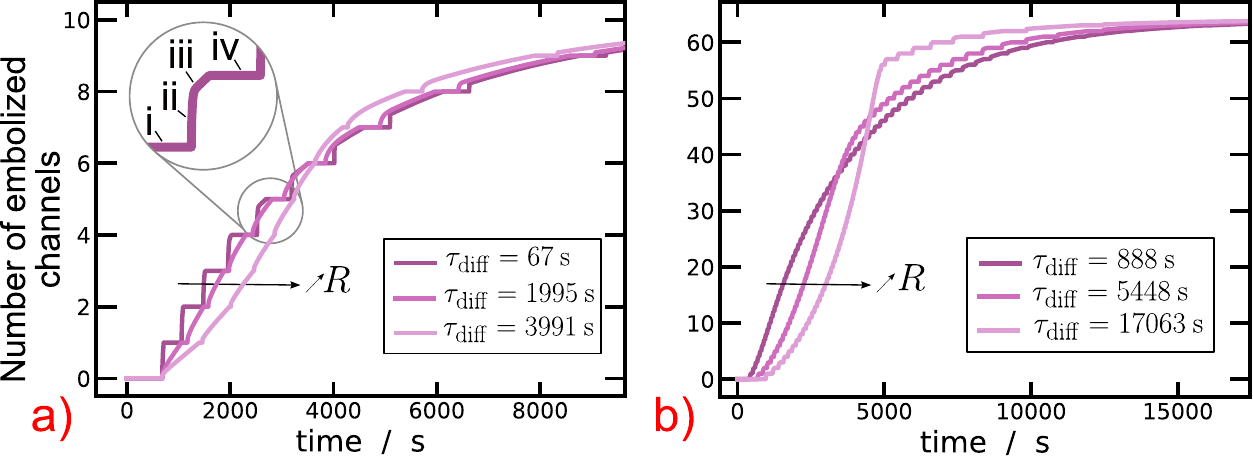}
    \caption{\textbf{Hydrodynamic resistance initially delays embolism front propagation, followed by a rapid "catch-up" dynamics.} Position of the front (channel index) versus time is plotted as hydrodynamic resistance $R$ is increased. For vanishing hydrodynamic resistance (dark magenta), the dynamics is that of a truncated exponential, as demonstrated by Dollet \textit{et al.} \cite{dollet_drying_2019}. As $R$ increases (lighter magenta), an inflection point appears in the shape of $L$ vs. $t$, a signature of the delay induced by the slow diffusion of the pressure in the resistive network. The values of the pressure diffusion timescale $\tau_\mathrm{diff} = N_0^2RC$ are expressed in the legend (see also Eq. \ref{eq:tau_diff}), and show that the inflection of the shapes appears as $\tau_\mathrm{diff}$ becomes comparable to the duration of the network pervaporation. (a) Example with 1D networks of $N_0=10$ channels and with $w=400\,\mu$m, $l=5\,$cm, $w_c = 15\,\mu$m and $l_c = 0.5$, $15$ or $50\,$mm. The embolism front advances intermittently, following the sequence of steps illustrated in Fig.~\ref{fig:serie}a. (b) Example given with 1D networks of $N_0=64$ channels (as in the following figures), with $w=400\,\mu$m, $l=0.5\,$cm, $w_c = 10$, $15$ or $30\,\mu$m and $l_c = 10\,$mm. See table \ref{tab:parameters} for the other parameters considered in this study.}
    \label{fig:emb_L_ex}
\end{figure*}

In fluidic networks as represented in Fig.~\ref{fig:serie}, the pressure in a channel is coupled to the one in the other channels by the constrictions that connect them and enable the fluid to flow between channels. The dynamics of the pressure ($p_i$) in the channel $i$ away from the embolism front is governed by the pressure in the neighboring channels and its pervaporation flux. One can write the following local water volume conservation equation for the channel $i$:
\begin{equation}
    \dv{t}p_i = \frac{(p_{i+1} - p_i) - (p_i - p_{i-1})}{RC} - \frac{jl}{C},
    \label{eq:gov_eq}
\end{equation}
which corresponds to the discretized version of a diffusion problem with a discrete diffusion coefficient $1/RC$ associated to a diffusion time $RC$. The first term on (\ref{eq:gov_eq})'s right-hand side accounts for pressure diffusion between neighboring channels, while the second represents the pressure drop induced by pervaporation. When the embolism front is stopped at a constriction, the pressure in the channel ($k$) next to the constriction at which the interface is stopped evolves as:
\begin{equation}
    \dv{t}p_k = \frac{p_{k+1} - p_k}{RC} - \frac{jl}{C}.
    \label{eq:gov_eq_k_stop}
\end{equation}
When the pressure $p_k$ in the channel $k$ reaches the Laplace pressure of the constriction, $p_c$, the interface front moves forward in the channel. We consider here that the mechanical relaxation of the channel section is instantaneous after the constriction jump leading to a reduction of the filled length of the channel to $\frac{V_k}{wh} = l + \frac{Cp_c}{wh}$. 
This assumption is consistent with the assumption made that the flow in constrictions is the dominant dissipative process. Then, the water volume $V_k$ in the draining channel decreases as:
\begin{equation}
    \dv{t}V_k = \frac{p_{k+1} - p_k}{R} - j\frac{V_k}{wh}.
    \label{eq:gov_eq_k_draining}
\end{equation}

We numerically solved  (\ref{eq:gov_eq})-(\ref{eq:gov_eq_k_draining}) to capture  the intermittent dynamics of the embolism propagating through the series of channels and constrictions. We first illustrate the effect of an increase of the hydrodynamic resistance on advancing front dynamics in Fig.~\ref{fig:emb_L_ex}. We can see the initial propagation of the embolism is greatly delayed when hydrodynamical resistance increases. This delay results from increased viscous dissipation in the network, which weakens the pressure gradient pulling the interface forward.

\begin{figure*}[t]
    \centering
    \includegraphics[width=0.9\textwidth]{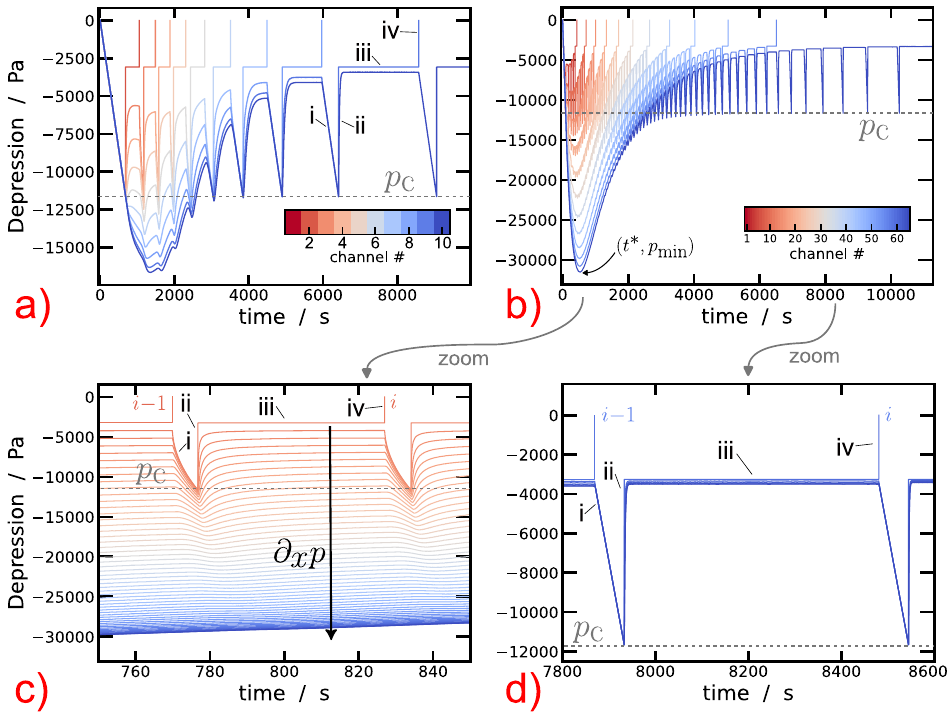} \caption{\textbf{Pressure dynamics in the channel series}. (a) Pressure evolution in successive channels (color-coded) as a function of time, for $N_0=10$. (b) Depression in the successive channels, coded by color, as a function of time, for $N_0=64$. Only a subset of $12$ channels are shown here. (c) Zoom on the pressure minimum ($p_\mathrm{min}$ at time $t^*$) observed in (b), where the series has significant hydrodynamic resistance. (d) Pressure dynamics near a single embolism event (steps $\mathsf{i}$–$\mathsf{iv}$ in panel a of Fig.~\ref{fig:serie}).}
\label{fig:emb_p_ex}
\end{figure*}

Fig.~\ref{fig:emb_p_ex}, shows the typical evolution of the pressure with time in the different channels of a relatively short ($N_0 = 10$, panel a)) and relatively long ($N_0=64$, panel b)) series. In this figure, we illustrate the successive steps of the pressure dynamics in the networks as the embolism front stops at a constriction and then progresses in the successive channel. Pressure dynamics in the channels exhibit pressure dips associated with the passage of the embolism front through the constrictions. On top of this impulse train dynamics, a transient pressure minimum is observed associated with significant viscous dissipation occurring in the system.

\subsection{A continuum model for the embolism spreading}
\label{subsec:continuous}

We now derive a continuous description of embolism propagation from the discrete model, describing the propagation of the liquid length $L(t)$ in the $x$ direction, as well as the pressure field $p(x,t)$. The continuum description is valid when considering long series and looking at their large size and long time dynamics, that is when :
\begin{equation}
    \left.\begin{array}{c}
         L_0 = N_0l \gg l
    \end{array}\right. 
    \qquad \& \qquad 
    \left(\begin{array}{c}
         \dd L, \dd x \gg l  \\
         \dd t \gg \Delta t_\mathrm{tot} 
    \end{array},\right. 
\end{equation}
where $L_0$, the cumulated length of all the channels, is a finite quantity and $\Delta t_\mathrm{tot}$ corresponds to the time that the embolism front takes to pass through a constriction and its following channel.
Following this procedure, equation (\ref{eq:gov_eq}) becomes:
\begin{equation}
    \partial_tp = D\partial_{x}^2 p - \frac{j}{c},
    \label{eq:diflike}
\end{equation}
where $D=\frac{1}{\mathcal{r}\mathcal{c}}$ with $\mathcal{r} = R/l$ and $\mathcal{c} = C/l$. This equation is a diffusion equation with a sink term ($-j/\mathcal{c}$). Concerning the boundary conditions, one have: in $x=L_0$, at the end of the channel, $\partial_x p = 0$, since no flow occurs at the closed end of the network; and, in $x = L_0-L$, at the interface, $p = p_\mathrm{c} = \gamma \kappa_\mathrm{c}$ \cite{NoteOnPc}. Eq. \ref{eq:diflike} is valid in the liquid-filled domain, whose size decreases as much as the liquid pervaporates and the embolism propagates. Thereby, it is more convenient to rewrite the equation in a fixed domain by doing the following appropriate change of spatial variable $\tilde{x} = 1-\frac{L_0-x}{L}$. We then find :
\begin{equation}
    \partial_t p + \frac{1-\tilde{x}}{L}\dot{L}\partial_{\tilde{x}}p = \frac{D}{L^2}\partial_{\tilde{x}}^2p - \frac{j}{\mathcal{c}},
    \label{eq:diff_fixDom}
\end{equation}
which holds between $\tilde{x}=0$  (corresponding to $x=L_0-L$, and where the boundary condition $p=p_\mathrm{c}$ is imposed) and $\tilde{x}=1$ (corresponding to $x=L_0$ and where the no-flux boundary condition $\partial_{\tilde{x}}p$ is imposed). Eq. (\ref{eq:diff_fixDom}) corresponds to the diffusion equation (\ref{eq:diflike}) rewritten in the Lagrangian interface reference frame. This variable change introduces an advection-like term: $(1-\tilde{x})/L\cdot\dot{L}\partial_{\tilde{x}}p$.

The system of equations can be closed by expressing volume conservation at the interface:
\begin{equation}
    \mathrm{r} \dot{L} S_\mathrm{c} = \partial_x p (L_0-L, t) ,
    \label{eq:VCons_local}
\end{equation}
with $S_\mathrm{c} = S_0 + Cp_\mathrm{c}/l$ the section of the channel when the pressure equals the Laplace pressure of the constriction (the condition to jump the constriction). From integration of (\ref{eq:diflike}), one obtains the following equation for $L$:
\begin{equation}
    S_\mathrm{c}\dot{L} + \qty(j + \overline{\partial_t S})L = 0,
    \label{eq:diff_fixDom_front}
\end{equation}
with $\overline{\partial_t S} = \mathcal{c}\int_{0}^{1}\partial_tp(\tilde{x},t)d\tilde{x}$. Note that Eq. \ref{eq:diff_fixDom_front} is equivalent to the global volume conservation condition $\dot{V}=-jL$ (see Appendix A).

\subsection{Comparison with non-deforming systems}

We now want to to more tightly relate our continuous model given by equations (\ref{eq:diff_fixDom}) \& (\ref{eq:diff_fixDom_front}) with the simpler model of drying channels having negligible compliance hydrodynamic resistance. Let first note that, even if $\overline{\partial_t S}$ does depend on $L$ (via $p$), one can integrate this equation for $L$ as if $\overline{\partial_t S}$ was a prescribed function of $t$. Thereby, one gets:
\begin{align}
    L &= L_0\exp(-\int_0^t \frac{j+\overline{\partial_t S}}{S_\mathrm{c}}\dd t') \notag \\
      &= L_0 \exp(-\frac{t}{\tau_\mathrm{pv}}) \exp(\frac{1}{S_c}\int^1_0 \mathcal{c}\qty[p_c - p(\tilde{x},t)]\dd \tilde{x}),
    \label{eq:L}
\end{align}
where we introduced $\tau_\mathrm{pv}$, the characteristic pervaporation time of the channel series:
\begin{equation}
    \tau_\mathrm{pv} = \frac{S_\mathrm{c}}{j}.
\end{equation}
This equation directly matches with the case of infinitely rigid channels with negligible viscous dissipation reported by Dollet \textit{et al.} \cite{dollet_drying_2019}: 
\begin{equation}
    L(t)\simeq L_0e^{-t/\tau_0},\:\mathrm{with }\: \tau_0 = S_0/j,
    \label{eq:L_Dollet}
\end{equation}
in the limit of negligible meniscus evaporation. In the latter case, the interface progression is straightforwardly forced by pervaporation in the absence of compliance of the network. For compliant systems, looking at the right hand side of equation (\ref{eq:L}), one can see that, in the case where all the volume lost by pervaporation is translated into channel depressurization, that is when $\mathcal{c}\qty(p_c-p(\tilde{x},t))=jt$, the interface front keeps static and one has $L=L_0$. Thus, in our configuration, the channel section modification delays the effect of pervaporation on the retraction of the interface, as the water loss is compensated by depressurization and section reduction. Conversely, when the section relaxes, it pulls on the interface. This effect renders interface dynamics more intricate than in the rigid case \cite{dollet_drying_2019, encarnacion_pervaporation-induced_2021, dollet_drying_2021}.

\subsection{Influence of the network design on the embolism dynamics}
\label{subsec:design}
\begin{figure*}[t!]
    \centering
    \includegraphics[width=0.8\textwidth]{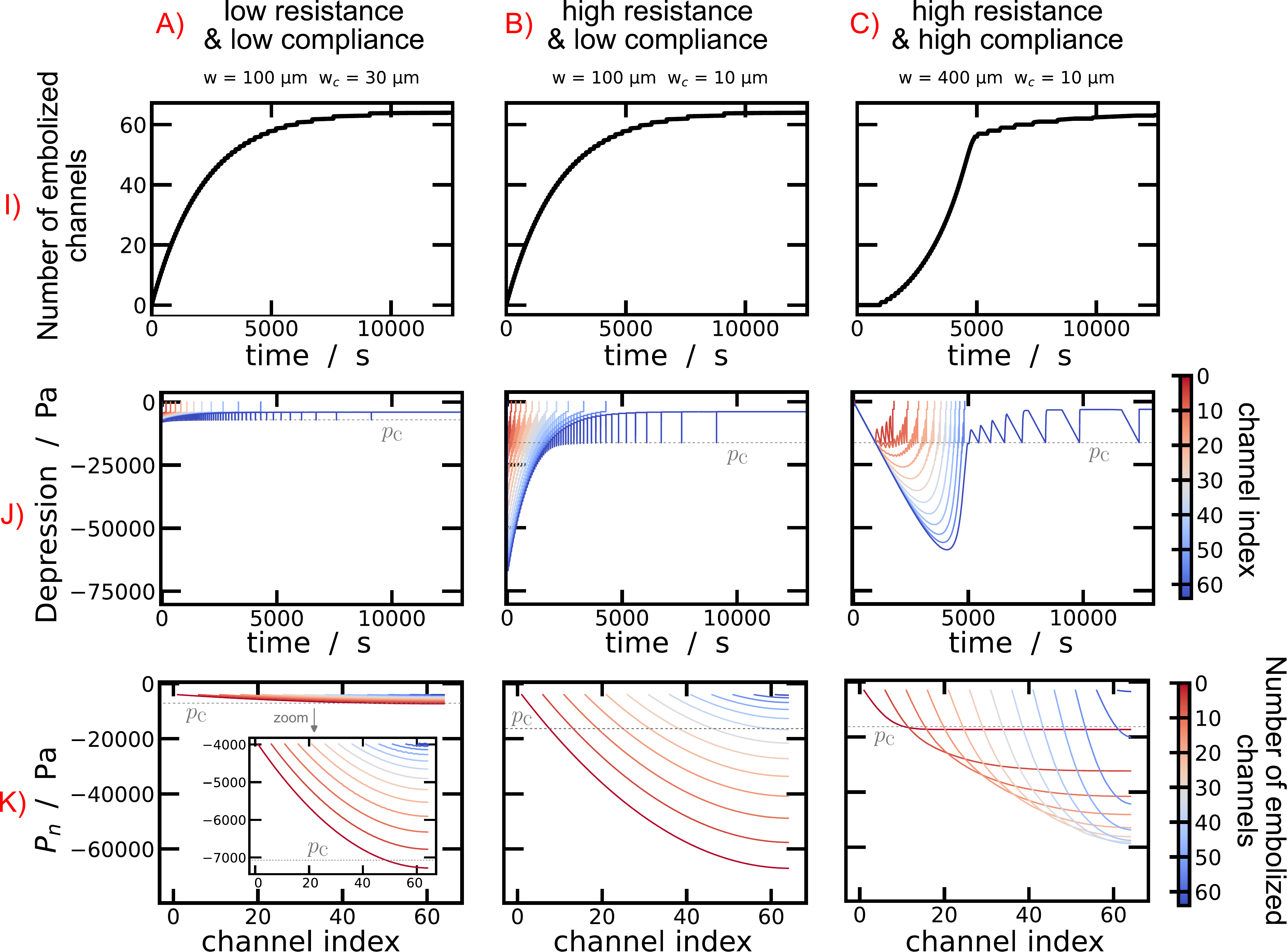}
    \caption{\textbf{Parametric analysis of the discrete model.} Resolution of the equations (\ref{eq:gov_eq}), (\ref{eq:gov_eq_k_stop}) and (\ref{eq:gov_eq_k_draining}) varying the design parameters of the series. First panel line represent the temporal evolution of the embolism front. Second panel line plot the dynamics of the pressure in the channels. Last panel line represent the pressure profile $P_n$ in the series where $P_n$ is the pressure at the channel $n$ when the embolism front just finished the draining of a channel. The series considered here have $N_0 = 64$, $l_c = 1\,$cm, $l = 5\,$mm and $h = 50\,\mu$m.}
    \label{fig:emb_param}
\end{figure*}

We conducted a parametric analysis of the discrete model described by Eqs. \ref{eq:gov_eq}, \ref{eq:gov_eq_k_stop} and \ref{eq:gov_eq_k_draining}, varying the width of the channels $w$ and the width of the constrictions $w_c$. The main outcomes of this parametric analysis is summarized Fig.~\ref{fig:emb_param}, where three archetypal cases are presented. For each column, the successive rows present respectively, the embolism front trajectory (first row (I)), the channel depressurization dynamics (second row (J)) and the evolution of the pressure $P_n$ (third row (K)). The first column (A) of Fig.~\ref{fig:emb_param} shows the results for the case of a large constriction width ($w_c = 30\,\mu$m) and small channel width ($w = 100\,\mu$m). This design implies a low hydrodynamic resistance $R$ coupled with a low compliance $C$ of the channel. This case is analogous to the one described in a recent study \cite{keiser_embolism_2024}, where pressure tends to homogenize rapidly after each jump of the interface that passes a constriction. The second column (B) of Fig.~\ref{fig:emb_param} combines small compliance ($w=100\,\mu$m) with large hydrodynamic resistance ($w_c = 10\,\mu$m). A large pressure gradient appears quasi-instantaneously, with a pressure minimum in the terminal channel that largely exceeds the capillary pressure of the constrictions. However, the trajectory of the embolism front in the series is not affected, such that cases A and B exhibit the same exponential dynamics. Moreover, row (K) shows that case (A) and (B) have similar pressure profiles that all seems parabolic. Finally, column (C) of Fig.~\ref{fig:emb_param} corresponds to a network with a large compliance ($w=400\,\mu$m) coupled to a large hydrodynamic resistance ($w_c = 10\,\mu$m). This design exhibits a large diffusion timescale $RC$ for the pressure. In this case, the trajectory of the embolism front (Fig.~\ref{fig:emb_param}.IC) is significantly modified compared to previous cases. In Fig.~\ref{fig:emb_param}.JC, a large depressurization peak appears in the channel, as in Fig.~\ref{fig:emb_param}.JB, but only after a significant delay corresponding to about one third of the total pervaporation time of the network. In Fig.~\ref{fig:emb_param}.KC, the pressure profile appears no more parabolic, a consequence of the strong coupling between the pressure field and the embolism front dynamics. Indeed, initially, the pressure profile is completely flat over a significant length away from the embolism front. As the embolism front advances the pressure gradient expands further over the network. It is only when the pressure gradient expanded towards the terminal channel that the latter starts to re-pressurize. The pressure typically diffuses on the length $N_0$ of the series with a characteristic time $\tau_\mathrm{diff}$ so that:
\begin{equation}
    \tau_\mathrm{diff} = N_0^2RC = \frac{L_0^2}{D}.
\label{eq:tau_diff}
\end{equation}
Overall, the parametric analysis showed the strong influence of the hydrodynamic resistance $R$ and the compliance $C$ on the onset of nonlinear dynamics and the appearance of a strong pressure minimum.

\begin{figure*}[t]
    \centering
    \includegraphics[width=0.75\linewidth]{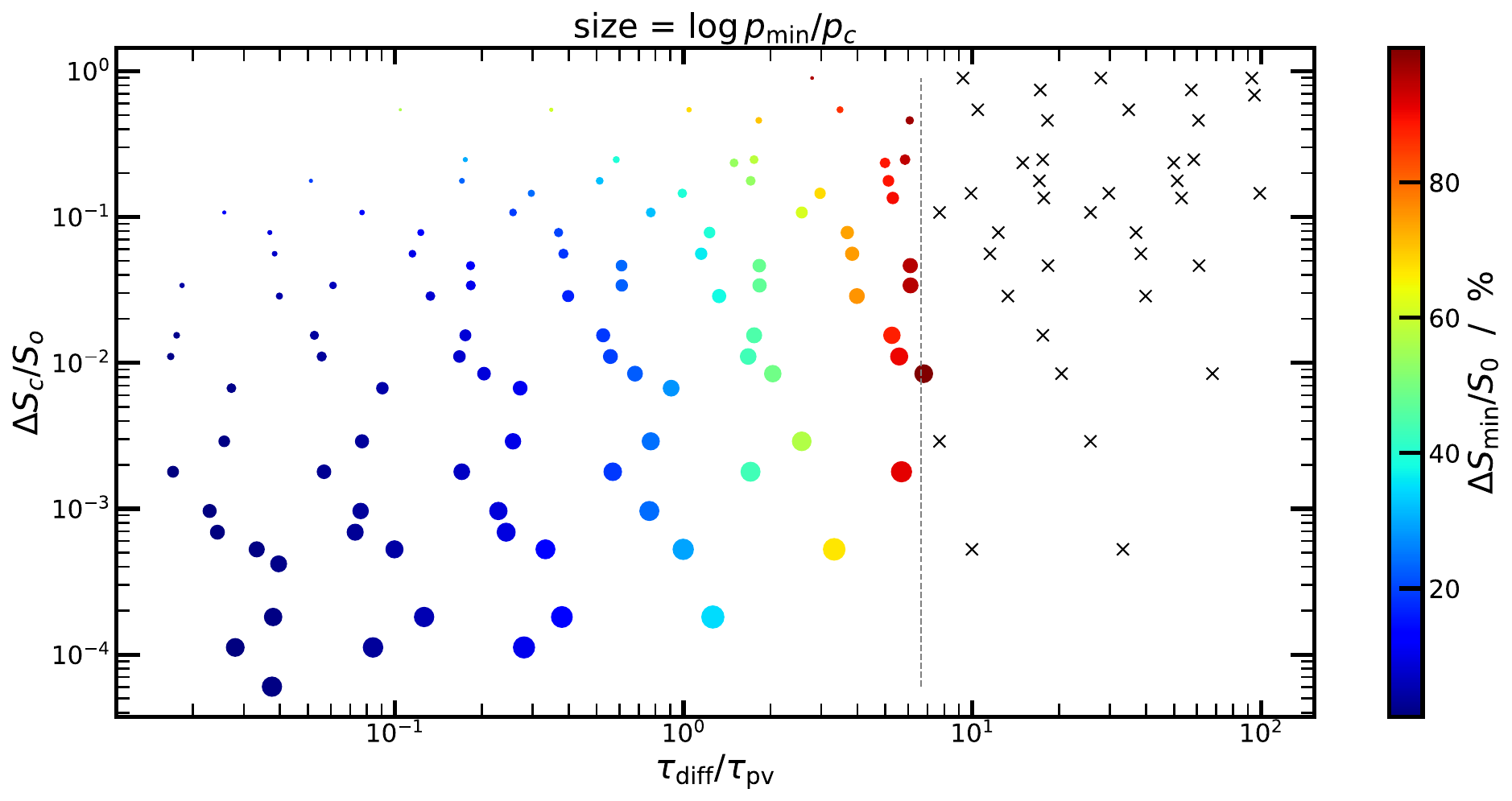}
    \caption{Parametric analysis of the continuum model of equations (\ref{eq:diff_fixDom}) and (\ref{eq:diff_fixDom_front}). The different design parameters of the series considered are plotted in the non-dimensional phase space ($\tau_\mathrm{diff}/\tau_\mathrm{pv}$, $\Delta S_\mathrm{c} / S_0$). Color and size of the scatter dots represent, respectively, the maximal deformation of the terminal channel ($\Delta S_\mathrm{min} / S_0$) and their minimal pressure ($p_\mathrm{min}/p_c$, log scale) over the course of the embolism propagation. Crosses indicate cases where the terminal channel undergoes full collapse owing to insufficient pressure compensation. The data correspond to a wide variety of design parameters with $w \in [50, 600]\ \mu$m, $l_c \in [0.3; 10^3]\ $mm and $w_c \in [1; 100]\ \mu$m.}
    \label{fig:phase_diag}
\end{figure*}

As we considered here long channel series ($N_0=64$), the main features observed from the discrete model (intricate interface dynamics, pressure minimum and pressure gradient profiles) can be reproduced from the continuous model of equations (\ref{eq:diff_fixDom}) and (\ref{eq:diff_fixDom_front}). We conducted a parametric sweep of the continuous model varying several geometric design parameters, whose output is represented in Fig.~ \ref{fig:phase_diag} as a phase diagram. The nondimensionalized phase parameters used in the diagram are $\tau_\mathrm{diff}/\tau_\mathrm{pv}$, which increases when pressure diffusion delay increases, and $\Delta S_\mathrm{c} / S_0$, which characterizes the deformation of the channels owing to the capillary pressure of the constrictions \cite{continuous}. Note that the continuous model was solved using a simplified version of the pervaporation rate, only keeping the linear dependence of $j$ with the channel width and neglecting the corrective terms of Eq. \ref{eq:pervap}. The maximal deformation of the terminal channel, $\Delta S_\mathrm{min}/S_0$, is observed to increase with $\tau_\mathrm{diff}/\tau_\mathrm{pv}$ (color shifting from blue to red). The ratio $p_\mathrm{min}/p_c$ of the pressure minimum and the capillary pressure of the constriction is observed to decrease (marker size reducing) as $\Delta S_\mathrm{c} / S_0$ increases. It is also noteworthy that for a ratio $\tau_\mathrm{diff}/\tau_\mathrm{pv}$ larger than one, the depressurization of the terminal channel is so intense that a total collapse of the terminal channel is observed, with outcomes falling outside the validity range of the current model. Channel collapses are marked with on the diagram.

In the following section, we explore how to rationalize analytically the results obtained from the parametric analysis. In particular, we look for prediction of the amplitude and time of appearance of the pressure minimum (respectively, $p_\mathrm{min}$ and $t^\ast$), as a function of the design parameters of the fluid network.

\section{Analytical developments}
\label{sec:discussion}

In the previous part, we demonstrated that the collective behavior of pervaporating channels can give rise to complex, nonlinear dynamics.  These nonlinear patterns are not only observed at the interface, through constriction-induced jumps \cite{keiser_embolism_2024}, but also in the global pressure distribution, which becomes inhomogeneous when the hydrodynamic resistance $R$ and compliance $C$ are large. Two main regimes can be distinguished depending on the ratio of the pressure diffusion timescale $\tau_\mathrm{diff}$ to the pervaporation timescale $\tau_\mathrm{pv}$. In the following, we provide analytical solutions for the pressure dynamics for the first case ($\tau_\mathrm{diff}\ll\tau_\mathrm{pv}$). For the second case ($\tau_\mathrm{diff}\sim\tau_\mathrm{pv}$), we also provide analytical estimates. We do not consider here the regime $\tau_\mathrm{diff} \gg \tau_\mathrm{pv}$, as it leads to uncontrolled channel collapse, where pressure relaxation is too slow to compensate for pervaporation-driven volume loss. Overall, the aim of this section is to rationalize the dynamics to expect for any chosen design to provide predictive guidelines for the design and analysis of both artificial and biological vascular networks.

\subsection{Networks with rapid pressure diffusion: $\tau_\mathrm{diff} \ll \tau_\mathrm{pv}$}
\label{subsec:larger}
\subsubsection{Discrete Model}

For fully coherent networks where pressure relaxation diffuses instantaneously on the entire network, the discrete model of equations (\ref{eq:gov_eq}), (\ref{eq:gov_eq_k_stop}) and (\ref{eq:gov_eq_k_draining}) can be solved analytically. Indeed, one can consider that, after a constriction jump and a quick transient time, the water flux and the pressure field reach a (temporary) stationary state during the drainage of the channel. This stationary state imposes that pressure suction at each constriction equals the pervaporative loss of the upward channels. Thus, during the stationary state of the drainage, one has:
\begin{equation}
    \frac{p_i-p_{i+1}}{R} = (N_0-i)jl
    \label{eq:stat_grad_i}
\end{equation}
In particular, this equation shows that pressure difference between successive channels should decrease with $i$ as the cumulated pervaporation flux of the upward channels decreases. Notably, equation (\ref{eq:stat_grad_i}) gives that the pressure gradient $\frac{p_i-p_{i+1}}{l}=(N_0-i)jR$, at a given channel position $i$ is independent of time for a given network. The speed of the interface, $v = \frac{l}{\Delta t_\mathrm{drain}} = \frac{(N_0-k)jl}{hw} = \frac{p_k-p_{k+1}}{Rhw}$ noting $k$ the index of the draining channel, is itself not stationary; it decreases, as well as the pressure gradient near the interface $\frac{p_k-p_{k+1}}{l}$, when the interface moves from a channel to the next (as $k$ increases).

Equation (\ref{eq:stat_grad_i}) can be integrated to yield $p_i$ explicitly:
\begin{equation}
    p_i = -\gamma \kappa-jlR\left(N_0 - \frac{i+k-1}{2}\right)(i-k)
    \label{eq:stat_grad_i_sol}
\end{equation}
Here $-\gamma \kappa$ is the residual Laplace pressure of the channel $k$ that is draining (see eq. \ref{eq:Laplace_p_c} considering the dimensions $h$ and $w$ of the channels). Equation (\ref{eq:stat_grad_i_sol}) predicts that the pressure profile of Fig.~\ref{fig:emb_param}.KA and KB is indeed parabolic. Noting $n = i-k$, the index of channels relative to the index $k$ at which the interface advanced, we can rewrite the equation as $\frac{p_i+\gamma \kappa}{jlR} = (N_0-k)n-(n-1)n/2$. Thus, in rescaled pressure units, curves of Fig.~\ref{fig:emb_param}.KA and KB indeed only depends on the remaining liquid-filled length, and no other parameters of the network.

Finally, one can notice that considering series without pressure relaxation delay ($\tau_\mathrm{diff}\ll\tau_\mathrm{pv}$) is equivalent to considering small number of channels $N_0$, so that during draining all channels relaxes in response to the advancement of the interface. $N_0$ has thus to be small compared to the number of channels over which the pressure diffuses during drainage $\sqrt{\frac{1}{RC} \Delta t_\mathrm{drain}} = \sqrt{\frac{1}{N_0}\frac{hw/j}{R C}}$ where $ hw/j$ is the characteristic time of pervaporation of an individual channel and $RC$ is the characteristic time of diffusion of the pressure between two channels. This is typically fulfilled when $N_0^2 \ll N^{*^2} = \frac{hw/j}{R C}$, that is, when $\tau_\mathrm{diff} \ll \tau_\mathrm{pv}$. We can thus express the characteristic number of channels $N^*$ for which the pressure relaxation delay starts versus the geometric parameters of the series. Assuming $w_\mathrm{c} < h_\mathrm{c} = h$ (one has to inverse the order of $w_\mathrm{c}$ \& $h_\mathrm{c}$ in the following expression otherwise) and considering only the first order dependence over the geometric parameters, one gets:   
\begin{equation}
    N^* \propto \sqrt{\frac{h\delta^4 w_\mathrm{c}^3 h_\mathrm{c}}{l_\mathrm{c}lw^5}}.
    \label{eq:N_star}
\end{equation}
$N^*$ can be considered as the critical network size, above which pressure relaxation cannot keep pace with pervaporation, so that nonlinear effects emerge. This equation indicates how the design parameters of the network ($N_0$, $l$, $w$, $h$, $l_\mathrm{c}$, $w_\mathrm{c}$, $h_\mathrm{c}$ and $\delta$) should be selected in order to observe similar embolism dynamics.

One can similarly express the maximal depression in the terminal channel in leading order of the design parameters with: $jlR\frac{N_0(N_0+1)}{2} \propto \frac{w l l_\mathrm{c}}{\delta w_\mathrm{c}^3 h_\mathrm{c}}\frac{N_0(N_0+1)}{2}$. Therefore, assuming $w \gg w_\mathrm{c}$, the ratio between the minimal pressure in the terminal channel $p_\mathrm{min}$ and the Laplace pressure imposed by the constriction goes as
\begin{equation}
\frac{p_\mathrm{min}}{p_c} \propto \frac{w l l_\mathrm{c}}{\delta w_\mathrm{c}^2 h_\mathrm{c}}\frac{N_0(N_0+1)}{2}.
\end{equation}
Assuming a stationary pressure profile during drainage of the channels leads to an analytical solution for the pressure profile in the series. This enables to rationalize the depressurization peak observed at the terminal channel of series without pressure diffusion delay in the middle column of Fig.~\ref{fig:emb_param} and to relate it to the series design parameters. In this simplified case, the pressure gradient observed is merely the suction profile needed to sustain the pervaporation flux, given the viscous dissipation in the series.

\subsubsection{Continuum limit}
\label{sec:noDiff_contLim}
In the continuous description, the absence of delay in the diffusion of the pressure relaxation implies that pressure gradient always drains as much water volume as it is lost by pervaporation, i.e. that: 
\begin{equation}
    -\frac{1}{\mathcal{r}}\partial_x p(x,t) = j(L_0-x),
    \label{eq:stat_grad}
\end{equation}
which is the continuous equivalent of equation (\ref{eq:stat_grad_i}). Combined with the volume conservation condition (\ref{eq:VCons_local}), it integrates into, $L = L_0e^{-t/\tau_\mathrm{pv}}$ for the liquid length (i.e. finding back the exponential decay from Dollet \textit{et al.} \cite{dollet_drying_2019}), and for the pressure field:
\begin{align}
    p(x,t) &= p_\mathrm{c} - \mathcal{r}j\left(x - (L_0 - L)\right)\frac{L_0+L-x}{2} \notag \\
           &= p_\mathrm{c} - \mathcal{r}j\left(x - L_0(1 - e^{-t/\tau_\mathrm{pv}})\right)\frac{L_0(1+e^{-t/\tau_\mathrm{pv}})-x}{2},
    \label{eq:p_qs}
\end{align}
which can be considered as the continuum version of equation (\ref{eq:stat_grad_i_sol}).

The absence of pressure diffusion delay implies a clear timescale separation between the evolution of $L$ (having the characteristic timescale $\tau_\mathrm{pv}$) and $p$ (transiently adapting to any change in L with characteristic timescale $\tau_\mathrm{diff}$). Here, $p$ is a fast variable (``slaved variable"), instantaneously adapting to the slow evolution of $L$. In fact, equation (\ref{eq:stat_grad}) can be obtained from integrating (\ref{eq:diff_fixDom}), supposing $p$ stationary and $L$ constant so that the evolution of the pressure field is quasi-static (quasi-stationary) and overall evolves with the characteristic time $\tau_\mathrm{pv}$.

\subsection{Networks with large pressure diffusion delay  $\tau_\mathrm{diff} \sim \tau_\mathrm{pv}$}
\label{subsec:smaller}

We now turn to the opposite regime, where pressure diffusion becomes comparable to pervaporation, leading to a delayed depressurization and a stronger coupling between pressure and embolism front dynamics.

\subsubsection{pressure minimum}

For networks presenting large pressure relaxation delays, the progression of the embolism front in the channel series owing to pervaporation proceeds over similar timescales as pressure diffusion in the network. Therefore, the embolism dynamics proceeds from a tight coupling between these two effects; such that both diffusion and interface advection terms play a role in the dynamics of the pressure field. These processes have respectively a typical timescale $\tau_\mathrm{diff}$ and $\tau_\mathrm{pv}$, and pressure in the network typically evolves with a timescale given by the composition of both processes:
\begin{equation}
\tau^\ast = \frac{\tau_\mathrm{pv}\tau_\mathrm{diff}}{\tau_\mathrm{pv}+\tau_\mathrm{diff}}. 
\end{equation}

\begin{figure*}[t]
    \centering
    \includegraphics[width=0.75\linewidth]{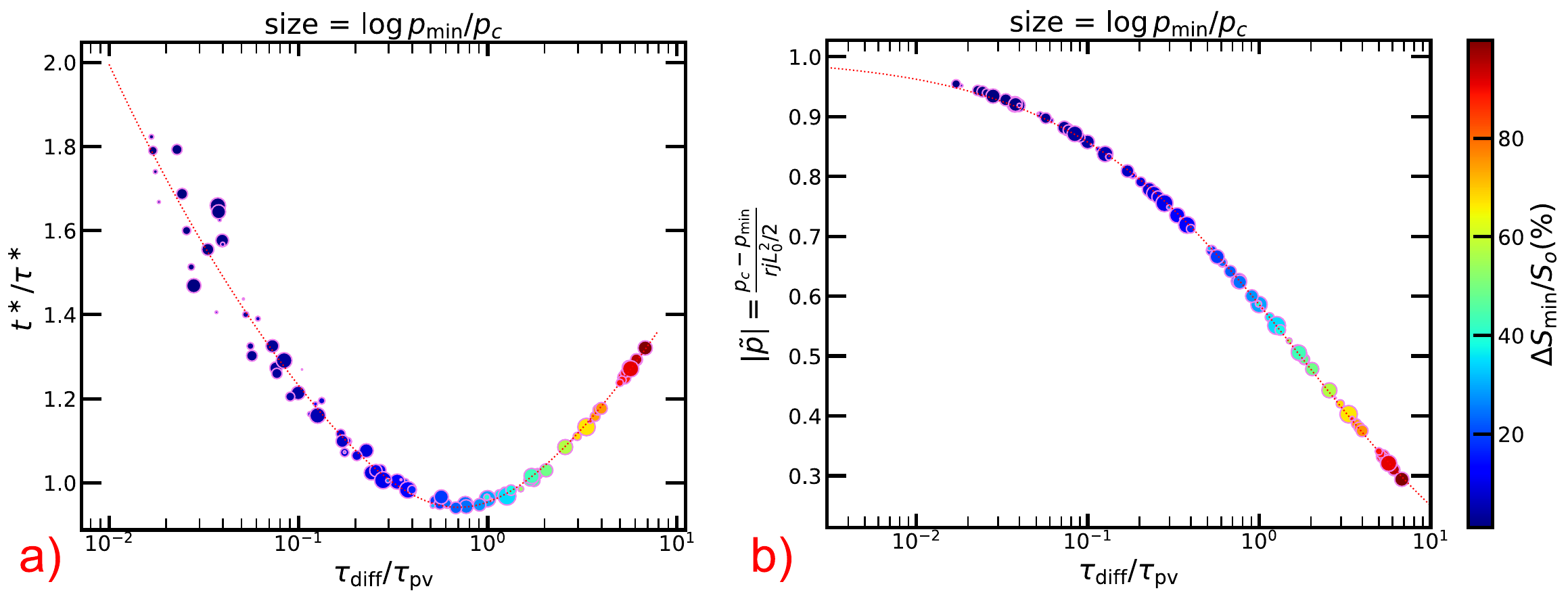}
    \caption{Collapse of the pressure minimum values is obtained by quasi-static and coupled rate process normalisation. (a) Time at which the pressure minimum is reached in the channel series $t^\ast$ normalized by $\tau^\ast = \frac{\tau_\mathrm{pv}\tau_\mathrm{diff}}{\tau_\mathrm{pv}+\tau_\mathrm{diff}}$ as a function of the ratio of the diffusion and the pervaporation time. The parametric exploration of the continuous model confirms that $t^\ast$ is close to $\tau^\ast$ over a large range of $\frac{\tau_\mathrm{diff}}{\tau_\mathrm{pv}}$ values. The red dotted-line is provided as a guide for the eyes, its equation is $\frac{t^\ast}{\tau^\ast}= 0.88\left(\frac{0.49+\left({\tau_\mathrm{diff}}/{\tau_\mathrm{pv}}\right)^2}{{\tau_\mathrm{diff}}/{\tau_\mathrm{pv}}}\right)^{0.21}$. (b) Nondimensionalized pressure minimum of the channel series versus $\tau_\mathrm{pv}/\tau_\mathrm{diff}$. The dashed red fitting curve is given by the equation $f\left(\tau_\mathrm{diff}  \tau_\mathrm{pv}\right) = 1/\left(1+0.7\left(\tau_\mathrm{diff} / \tau_\mathrm{pv}\right)^{0.63}\right)$.}
    \label{fig:cont_model_P}
\end{figure*}

In the early times of the embolism spreading, the gradient profile spreads slowly over the network. The pressure and the section of terminal channels only decrease owing to the volume they lose by pervaporation, regardless of the movement of the interface that does not influence them. A minimum for the pressure $p_\mathrm{min}$ is then reached at $t^\ast \sim \tau^\ast$, when a significant number of channels has been embolized and the relaxation of the pressure has started to diffuse from the front to end of the channel. As shown in Fig.~ \ref{fig:cont_model_P}.a, $t^\ast$ is indeed found to be of the order of $\tau^\ast$, with $t^\ast$ varying between $\tau^\ast$ and $2\tau^\ast$ for $\tau_\mathrm{diff}/\tau_\mathrm{pv}$ varying from $10^{-2}$ to $8$. Note that, close to $\tau_\mathrm{diff}/\tau_\mathrm{pv} = 1$, we have $t^\ast = \tau^\ast$. Moreover, data for $t^\ast / \tau^\ast$ all collapse on a unique curve which varies symmetrically around $\tau_\mathrm{diff}/\tau_\mathrm{pv} = 1$.

All the values of the minimal pressure $p_\mathrm{min}$ for the different design parameters are also found to collapse onto a single curve when rescaling the depression $p_\mathrm{min}-p_\mathrm{c}$ by its value in the quasi-stationary regime $j\mathcal{r}L_0^2$ and looking at the dependence of this depression with the ratio $\tau_\mathrm{diff} / \tau_\mathrm{pv}$, as shown in Fig.~ \ref{fig:cont_model_P}.b. This collapse suggests that the nonlinear interplay between embolism front dynamics and pressure field evolution is fully captured by the dimensionless ratio $\tau_\mathrm{diff} / \tau_\mathrm{pv}$. Following this rescaling, noting $\tilde{p} = \frac{p-p_\mathrm{c}}{j\mathcal{r}L_0^2}$ and $\tilde{t} = t/\tau_\mathrm{pv}$, and further rescaling the liquid length as $\tilde{L}=L/L_0$, we can rewrite the governing equation (\ref{eq:diff_fixDom}) and (\ref{eq:VCons_local}) as:
\begin{equation}
    \partial_{\tilde{t}} \tilde{p} + \frac{1-\tilde{x}}{\tilde{L}}\dd_{\tilde{t}} \tilde{L}\partial_{\tilde{x}}\tilde{p} = \frac{\tau_\mathrm{pv}}{\tau_\mathrm{diff}}\frac{\partial_{\tilde{x}}^2 \tilde{p}}{\tilde{L}^2}- \frac{\tau_\mathrm{pv}}{\tau_\mathrm{diff}},
    \label{eq:diff_fixDom_ad}
\end{equation}
and:
\begin{equation}
    \dd_{\tilde{t}}\tilde{L} = \frac{\partial_{\tilde{x}}\tilde{p} (0, \tilde{t})}{\tilde{L}},
    \label{eq:VCons_local_ad}
\end{equation}
with initial conditions $\tilde{L}=1$ and $\tilde{p} = 0$ and boundary conditions $\tilde{p}(0,\tilde{t}) = 0$ and $\partial_{\tilde{x}}p(1,\tilde{t}) = 0$. Although Eq. (\ref{eq:diff_fixDom_ad}) is reminiscent of a linear one-dimensional convection-diffusion equation, the Lagrangian interface condition in Eq. (\ref{eq:VCons_local_ad}) renders the problem nonlinear and spatially nonlocal, because the advective velocity depends on the moving interface and on the global pressure field. In the limit $\tau_\mathrm{pv} \gg \tau_\mathrm{diff}$, as the advection term becomes negligible compared to the diffusion term, we recover the separation of variables discussed in paragraph \ref{sec:noDiff_contLim}, as well as parabolic pressure profiles. In particular, pressure rapidly relaxes to the quasistatic profile (\ref{eq:p_qs}), while $\tilde{L}$ slowly evolves as $\exp(-\tilde{t})$. Equations (\ref{eq:diff_fixDom_ad}) and (\ref{eq:VCons_local_ad}) show that any parametric exploration of the embolism dynamics can be reduced to variation of the unique parameter $\tau_\mathrm{diff}/\tau_\mathrm{pv}$ for the dimensionless variables $\tilde{p}$, $\tilde{L}$ and $\tilde{t}$. Fig.~\ref{fig:cont_model_P}.b shows the collapse of all the minimal pressure data of the parametric exploration previously present in Fig.~\ref{fig:phase_diag}. Thus, $\tilde{p}  = \frac{p-p_\mathrm{c}}{j\mathcal{r}L_0^2}$ is found to be a decreasing function of $\tau_\mathrm{diff}/\tau_\mathrm{pv}$. Interestingly, this non-dimensionalization may mask non-monotonic dependencies of $p_\mathrm{min}$ on geometric design parameters. For instance, $p_\mathrm{min}$ is non-monotonic with the width of the channels $w$ owing to the competitive effect of w on $\tau_\mathrm{diff}/\tau_\mathrm{pv} \propto w^5$ and on $j\mathcal{r}L_0^2 \propto w$.

\subsubsection{Embolism length dynamics}

\begin{figure*}[t]
    \centering
    \includegraphics[width=0.75\linewidth]{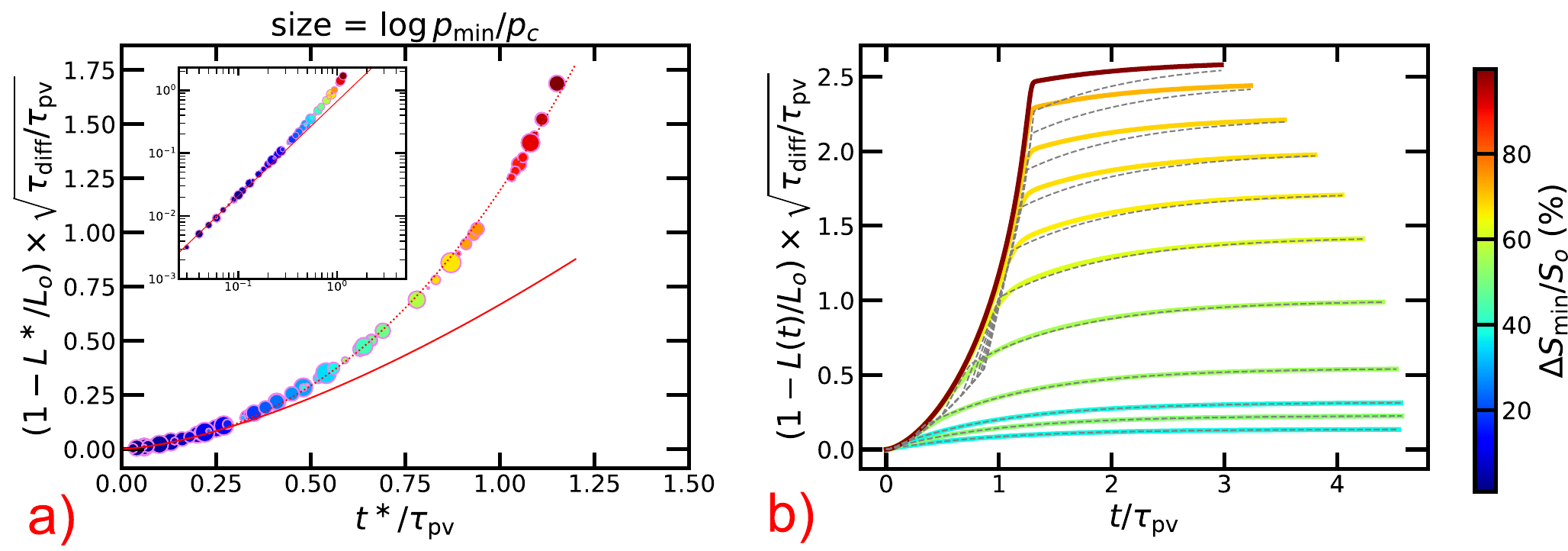}
    \caption{\textbf{Embolism dynamics of resistive and deforming channel series collapse on a piecewise master curve}. Non-dimensionalized embolism length is plotted as a function of time. In (a) all the data of Fig.~\ref{fig:phase_diag} are represented and length is plotted at $t^\ast$, the time at which the pressure minimum is reached. Inset is the log-log scale view. Red continuous curve has equation $f\left(\tau_\mathrm{pv}/\tau_\mathrm{diff}\right) =  2/3 \left(\tau_\mathrm{pv}/\tau_\mathrm{diff}\right)^{3/2}$.
    In (b) a representative selection of the length data is plotted. Red dashed lines plots the piecewise function of equation (\ref{eq:L_semi_anal}). This semi-analytical model appears to rather correctly fit the data.}
    \label{fig:cont_model_L}
\end{figure*}

We now try to predict the dynamics of the liquid length $L$. The speed of retraction of the interface is governed by the pressure gradient at the interface as outlined by equation (\ref{eq:VCons_local}). Slow diffusion confines the water injection, from the advancing of the interface, and pressure relaxation is confined near the interface, preventing the establishment of a stationary gradient over the full system. Absence of pressure relaxation on the terminal channels implies that all the water injected by the moving interface is removed from the system by the pervaporation over the length $\sim \sqrt{Dt}$ where the gradient establishes within time $t$. This hypothesis of limitation by the diffusion and the compensation of the water fluxes, only valid for $t\ll\tau_\mathrm{diff}\sim \tau_\mathrm{pv}$, can thus be translated into:
\begin{equation}
    S_\mathrm{c}\dot{L} \simeq -j\sqrt{Dt},
\end{equation}
which can be further solved into:
\begin{equation}
    L = L_0 - \frac{2}{3}\sqrt{D}\frac{j}{S_\mathrm{c}}t^{3/2},
    \label{eq:early_dyn_L}
\end{equation}
or:
\begin{equation}
    \left(1-\frac{L}{L_0}\right)\sqrt{\frac{\tau_\mathrm{diff}}{\tau_\mathrm{pv}}} = \frac{2}{3}\left(\frac{t}{\tau_\mathrm{pv}}\right)^{3/2}.
    \label{eq:early_dyn_L_2}
\end{equation}
Figures \ref{fig:cont_model_L}.a and \ref{fig:cont_model_L}.b show that this behavior for $L$ is indeed observed for $t < \tau_\mathrm{pv} / 4$ universally in all the trajectory obtained in the parametric exploration of the continuous model. For $t > \tau_\mathrm{pv} / 4$, $L(t)$ starts to deviate from this behavior. Indeed, the depression accumulated in the terminal channels starts to diffuse and pull on the interface so that the catching-up of pressure relaxation accelerates the movement of the interface. Surprisingly, during this phase, the embolism trajectories of the various series are still found to collapse in Fig.~\ref{fig:cont_model_L}.a and b. During this transition regime, the interplay between the accumulated depressurization in distal channels and the advancing interface results in an accelerating front, even though pervaporation remains constant. This regime is thus driven by delayed elastic recoil and restored pressure communication. Eventually, when the liquid length $L$ has been significantly reduced, the pressure diffusion over the remaining length become very fast and no more pressure relaxation delays affect the embolism dynamics, such that quasistatic regime for $p$ and exponential decreasing of $L$ are found back. Based on the previous observations, we build a semi-analytical solution for $L(t)$, smoothly transitioning between the diffusion-limited regime (eq. \ref{eq:early_dyn_L}) and the quasistatic regime (eq. \ref{eq:p_qs}), between $t=\tau^*$ and $t=1.5\tau^*$:
\begin{equation}
    L(t) = \left\{ 
    \begin{array}{lc}
         L_\mathrm{dl} = L_0 \left( 1 - \frac{2}{3}\frac{t^{3/2}}{\tau_\mathrm{pv}\sqrt{\tau_\mathrm{diff}}}\right) &  \text{for } t < \tau^*\\

         \sqrt{L_\mathrm{dl}^2 \left( 1 - \frac{t-\tau^*}{0.5\tau^*} \right) + L_\mathrm{qs}^2\frac{t-\tau^*}{0.5\tau^*}}& \text{for } \tau^* < t < 1.5\tau^* \\
                  L_\mathrm{qs} = L^+e^{-\frac{t-\tau^*}{\tau_\mathrm{pv}}}&  \text{for } t > 1.5\tau^*.
    \end{array}
    \right.
    \label{eq:L_semi_anal}
\end{equation}
Here, $L^+ \simeq L_0 \left(1 - \frac{\tau^*}{\tau_\mathrm{pv}} + \frac{4}{15}\frac{{\tau^*}^{5/2}}{\tau_\mathrm{pv}^2\tau_\mathrm{diff}} \right)$ is the liquid length ensuring volume conservation if the embolism dynamics were to switch instantaneously from a diffusion-limited regime ($L_\mathrm{dl}$) to a quasi-static regime ($L_\mathrm{qs}$) at $\tau^*$ while conserving the water volume.
As shown in Fig.~\ref{fig:cont_model_L}.b, this expression of $L(t)$ gives a rather correct prediction of the complex trajectory of the embolism depending on the design parameters of network. This analytical form provides a predictive tool for designing microfluidic systems with targeted embolism dynamics by tuning resistance and compliance.

\section{Perspectives}
\label{sec:perspectives}

\subsection{Fluidic networks}
\label{subsec:fluidic_networks}

Our results reveal how the interplay between compliance, viscous dissipation, and pervaporation can drive rich and nonlinear embolism dynamics in compliant microfluidic networks, dynamics which were previously unexplored.

Beyond embolism, nonlinear behaviors in related systems, such as memristor-like hydrodynamic resistance \cite{martinez-calvo_fluidic_2024} or fluidic Braess paradoxes \cite{case_braesss_2019}, suggest that pervaporating networks may also exhibit unexpected emergent phenomena. In our system, nonlinearity arises not from flow-rate-dependent resistance but from the evolving functional topology of the network, as the advancing embolism front dynamically alters hydraulic connectivity. This mechanism bears similarities to frangible porous media, where structural changes dynamically influence transport \cite{derr_flow-driven_2020, Zareei2022}. 

Extending this framework from one-dimensional to two-dimensional networks represents a promising next step, as branched and looped architectures are expected to introduce new collective behaviors and complex transport regimes. Previous investigations in networks with fast pressure diffusion evidenced embolism trajectories solely controlled by capillarity \cite{encarnacion_pervaporation-induced_2021,dollet_drying_2021}. In networks with slow pressure diffusion, our work suggests more subtle and intricate bubble propagation, and we can expect the induction of preferential embolism propagation paths, spatially localized collapse zones, or even front bifurcations, effects absent in one-dimensional networks. These investigations could be relevant both for fundamental studies of nonlinear dynamics \cite{fancher_mechanical_2022} and for applications in adaptive fluidic networks \cite{winn_operating_2024}.

On the experimental side, future studies could be guided by the design principles established here, leveraging recent pervaporation-driven methods \cite{bacchin_microfluidic_2022}. Previous PDMS biomimetic leaves were operated in the coherent regime \cite{keiser_embolism_2024, gauci2025channel}, with $\tau_{\mathrm{diff}}\ll\tau_{\mathrm{pv}}$ (Keiser \textit{et al.} \cite{keiser_embolism_2024}: $\tau_{\mathrm{diff}}\simeq 30$ s, $\tau_{\mathrm{pv}}\simeq 2000$ s). To access pressure-delay and the associated nonlinearities with similar microfabrication, one can increase the pressure-diffusion time $\tau_\mathrm{diff}=N_0^2RC$ by $10^2$–$10^3$ using the same soft-lithography toolbox: raise $R$ (narrower/longer constrictions), raise $C$ (softer/thinner walls, larger compliant chambers), and increase the number of repeated elements $N_0$. Such designs would generate front-coupled dynamics, while remaining compatible with pervaporation-driven actuation. Furthermore, experimental setups using PDMS microchannels under negative pressure offer promising avenues for validation \cite{bruning_turning_2019-1}. The possible triggering of cavitation events in regions far from the embolism front \cite{pingulkar_pervaporation-driven_2024} could significantly modify embolism progression patterns, opening new questions in the physics of drying fluidic systems, and in particular for plant-inspired networks.

\subsection{Embolism propagation in plants}
\label{subsec:plants}

While our model is designed as a minimal physical description of embolism dynamics in linear fluidic networks, it may offer useful analogies for interpreting embolism propagation in plant vascular systems, particularly in species with simple or weakly branched architectures, such as \textit{Adiantum} leaves or non-dividing stems \cite{brodribb_revealing_2016, keiser_embolism_2024, brodribb_optical_2017}. In natural xylem networks, additional processes, including lateral water storage and interactions with living tissues, further complicate the picture \cite{holtta_capacitive_2009}. Nonetheless, our model isolates a few key mechanisms (such as pressure diffusion delays and compliance-driven buffering) that could be relevant under certain conditions.  Embolism propagation in plants has been reported to occur intermittently and over a range of time scales \cite{brodribb_revealing_2016}, the origin of which remains unclear. In this context, pressure diffusion delays could contribute an additional dynamical ingredient. 

It has been documented that pressure diffusion in living xylem exhibits hydrodynamic delays set by hydraulic resistance–capacitance (RC) effects, typically from minutes to a few hours, in the absence of embolism \cite{phillips1997time, holtta_capacitive_2009, meinzer2004dynamics, blackman_two_2011, zhuang2014series, luo2021leaf}. In tropical canopy trees, storage-driven lags of order tens of minutes (up to about 1 h) are consistent with whole-tree time constants near 0.5 h \cite{meinzer2004dynamics}, and in loblolly pine diurnal lags of about 30–60 min were exhibited \cite{phillips1997time}. At the leaf scale of woody angiosperms, dynamic and bulk capacitances buffer changes in water potential on minute timescales \cite{blackman_two_2011}. Theoretical capacitive models formalize these trends with a characteristic time $\tau = RC$, predicting transients spanning tens to about 200 min for a monocot grass leaf \cite{luo2021leaf}. In plant-scale observations without embolism, the intermittency period is primarily paced by boundary forcing (e.g., evapotranspiration or root water potential) interacting with RC storage. By contrast, in our study it emerges endogenously from bordered-pit thresholding, where capillary constraints intermittently arrest propagation until pressure (tension) relaxes and re-establishes. Thus, our model provides a quantitative baseline for reinterpreting intermittent embolism sequences as delayed, RC-mediated pressure-diffusion phenomena.

While we model pit membranes as idealized constrictions with a capillary threshold, their actual structure is more complex. Intervessel pits in angiosperms contain porous and often deformable membranes \cite{zhang_gold_2024}, and gymnosperm pits include specialized torus-margo valves \cite{pittermann_torus-margo_2005, park_fluid-structure_2021}. These elements likely influence embolism propagation in ways not currently captured by our model. Yet, the idea that the pit geometry sets both a resistance and a capillary threshold remains qualitatively relevant.

Our findings suggest that large pressure drops may exceed capillary thresholds. Extending this approach to reticulate networks could help explore preferential embolism routes \cite{brodribb_optical_2017} or structural adaptations that maintain connectivity in vascular systems. Finally, the model opens the possibility that the parameter $\mathcal{r}L_0^2$, which governs pressure field evolution in the continuous limit, could be of functional significance. Whether such a quantity is constrained or tuned across species remains an open question, but it may offer a useful perspective for interpreting hydraulic behavior in terms of global network properties.

\section{Conclusion}
In this article, we theoretically investigated air invasion in one-dimensional networks of compliant microchannels undergoing pervaporation. While such systems were previously explored in our earlier work \cite{keiser_embolism_2024}, we extended the analysis by introducing a sufficiently large hydrodynamic resistance in the constrictions, which induces significant pressure and deformation inhomogeneities along the biomimetic veins. 

We identified two key timescales that govern the dynamics of this system: the intrinsic pervaporation timescale $\tau_\mathrm{pv}$ and the pressure diffusion timescale $\tau_\mathrm{diff}$. When these timescales become comparable (or when $\tau_\mathrm{diff} > \tau_\mathrm{pv}$) pressure within certain parts of the network, particularly near its distal end, can drop well below the capillary threshold $p_c$ required for embolism propagation.

By combining discrete network simulations and a continuum theoretical approach, we developed a predictive framework for the embolism dynamics. Within the continuous description, we showed that the system behavior could be fully captured by a single nondimensional parameter, $\tau_\mathrm{diff} / \tau_\mathrm{pv}$ (see Eqs. \ref{eq:diff_fixDom_ad} and \ref{eq:VCons_local_ad}). Furthermore, we demonstrated that a diffusion-limited approximation captures essential features of the nonlinear dynamics and provides a simple, semi-analytical understanding of interface propagation.

Overall, the system analyzed here offers a fundamental illustration of how nonlinear behavior can emerge in fluidic networks owing to the coupling between pressure field and interfacial motion. Our model provides a basis for further studies of nonlinear elasto-fluidic dynamics, both in the context of sap flow in plants and in engineered microfluidic systems. Extending this approach to branched or looped architectures, or to networks exhibiting instabilities and multistable flow paths, could reveal richer emergent phenomena.

 
\section*{Acknowledgement}
This work has been supported by the French government, through the grant ANR-19-CE30-0010 (PHYSAP) and through the UniCA Investments in the Future project managed by the National Research Agency (ANR) with the reference number ANR-15-IDEX-01. The authors want to thank Benjamin Dollet and Philippe Marmottant for fruitful discussions, as well as Virgile Thiévenaz, José Torres-Ruiz, Eric Badel and Hervé Cochard.

\section*{Data accessibility}

Numerical scripts and related data enabling to plot the figures are available in a Zenodo repository, accessible via https://doi.org/10.5281/zenodo.17485963, see Ref. \cite{zenodo17485963}. 

\newpage
\section*{Appendix A: global volume conservation condition in the continuum model}
\label{sec:appendixA}

We here demonstrate the equivalence of equation (\ref{eq:diff_fixDom_front}) with the global conservation of the water volume that must be fulfilled in the system. Noting $V$ the total volume of water in the system, global volume conservation can be expressed as follows :

\begin{equation}
\dot{V} = -jL
\end{equation}
Let first note that $V(t) = \int_{L_0-L}^{L_0}S(x, t)dx$ and $S(x, t) = S_0 + \mathcal{c} p(x, t)$, we have \begin{equation}
    V(t) = S_0 L(t) + \mathcal{c}\int_{L_0-L}^{L_0}p(x, t)dx
\end{equation}

To get the derivative of the volume, one has to use the Leibniz formula to account for the time-dependency of the integral boundaries. This gives :

\begin{align}
    \dot{V} &= S_0\dot{L} + c\dot{L_0}p(x=L_0, t) \notag \\
            &\quad - c\dv{t}(L_0-L)p(x=L_0-L, t) \notag \\
            &\quad + c\int_{L_0-L}^{L_0}\partial_tp(x,t)\,dx
\end{align}

As $L_0$ is a constant, the first term vanishes and noting that $\dv{t}(L_0-L) = -\dot{L}$, \ $p(x=L_0-L) = p_\mathrm{c}$\  and\  $S_c = \left(S_0 + \frac{Cp_\mathrm{c}}{l}\right)$, one can reexpress the global volume conservation as :

\begin{equation}
    \dot{V} = \dot{L}S_c + \mathcal{c}\int_{L_0-L}^{L_0}\partial_tp(x)dx = -jL
\end{equation}

A change of variables $\tilde{x} = 1 - \frac{L_0 - x}{L}$ in the integral enables to simplify the expression and to find back equation (\ref{eq:diff_fixDom_front}).


\newpage

\begin{thebibliography}{99}%
\makeatletter
\providecommand \@ifxundefined [1]{%
 \@ifx{#1\undefined}
}%
\providecommand \@ifnum [1]{%
 \ifnum #1\expandafter \@firstoftwo
 \else \expandafter \@secondoftwo
 \fi
}%
\providecommand \@ifx [1]{%
 \ifx #1\expandafter \@firstoftwo
 \else \expandafter \@secondoftwo
 \fi
}%
\providecommand \natexlab [1]{#1}%
\providecommand \enquote  [1]{``#1''}%
\providecommand \bibnamefont  [1]{#1}%
\providecommand \bibfnamefont [1]{#1}%
\providecommand \citenamefont [1]{#1}%
\providecommand \href@noop [0]{\@secondoftwo}%
\providecommand \href [0]{\begingroup \@sanitize@url \@href}%
\providecommand \@href[1]{\@@startlink{#1}\@@href}%
\providecommand \@@href[1]{\endgroup#1\@@endlink}%
\providecommand \@sanitize@url [0]{\catcode `\\12\catcode `\$12\catcode `\&12\catcode `\#12\catcode `\^12\catcode `\_12\catcode `\%12\relax}%
\providecommand \@@startlink[1]{}%
\providecommand \@@endlink[0]{}%
\providecommand \url  [0]{\begingroup\@sanitize@url \@url }%
\providecommand \@url [1]{\endgroup\@href {#1}{\urlprefix }}%
\providecommand \urlprefix  [0]{URL }%
\providecommand \Eprint [0]{\href }%
\providecommand \doibase [0]{https://doi.org/}%
\providecommand \selectlanguage [0]{\@gobble}%
\providecommand \bibinfo  [0]{\@secondoftwo}%
\providecommand \bibfield  [0]{\@secondoftwo}%
\providecommand \translation [1]{[#1]}%
\providecommand \BibitemOpen [0]{}%
\providecommand \bibitemStop [0]{}%
\providecommand \bibitemNoStop [0]{.\EOS\space}%
\providecommand \EOS [0]{\spacefactor3000\relax}%
\providecommand \BibitemShut  [1]{\csname bibitem#1\endcsname}%
\let\auto@bib@innerbib\@empty

\bibitem [{\citenamefont {Wootton}\ and\ \citenamefont {Ku}(1999)}]{wootton1999fluid}%
  \BibitemOpen
  \bibfield  {author} {\bibinfo {author} {\bibfnamefont {D.~M.}\ \bibnamefont {Wootton}}\ and\ \bibinfo {author} {\bibfnamefont {D.~N.}\ \bibnamefont {Ku}},\ }\bibfield  {title} {\bibinfo {title} {Fluid mechanics of vascular systems, diseases, and thrombosis},\ }\href@noop {} {\bibfield  {journal} {\bibinfo  {journal} {Annual review of biomedical engineering}\ }\textbf {\bibinfo {volume} {1}},\ \bibinfo {pages} {299} (\bibinfo {year} {1999})}\BibitemShut {NoStop}%
\bibitem [{\citenamefont {Marbach}\ \emph {et~al.}(2023)\citenamefont {Marbach}, \citenamefont {Ziethen}, \citenamefont {Bastin}, \citenamefont {Bäuerle},\ and\ \citenamefont {Alim}}]{marbach_vein_2023}%
  \BibitemOpen
  \bibfield  {author} {\bibinfo {author} {\bibfnamefont {S.}~\bibnamefont {Marbach}}, \bibinfo {author} {\bibfnamefont {N.}~\bibnamefont {Ziethen}}, \bibinfo {author} {\bibfnamefont {L.}~\bibnamefont {Bastin}}, \bibinfo {author} {\bibfnamefont {F.~K.}\ \bibnamefont {Bäuerle}},\ and\ \bibinfo {author} {\bibfnamefont {K.}~\bibnamefont {Alim}},\ }\bibfield  {title} {\bibinfo {title} {Vein fate determined by flow-based but time-delayed integration of network architecture},\ }\href {https://doi.org/10.7554/eLife.78100} {\bibfield  {journal} {\bibinfo  {journal} {eLife}\ }\textbf {\bibinfo {volume} {12}},\ \bibinfo {pages} {e78100} (\bibinfo {year} {2023})}\BibitemShut {NoStop}%
\bibitem [{\citenamefont {Le~Verge-Serandour}\ and\ \citenamefont {Alim}(2024)}]{le_verge-serandour_physarum_2024}%
  \BibitemOpen
  \bibfield  {author} {\bibinfo {author} {\bibfnamefont {M.}~\bibnamefont {Le~Verge-Serandour}}\ and\ \bibinfo {author} {\bibfnamefont {K.}~\bibnamefont {Alim}},\ }\bibfield  {title} {\bibinfo {title} {\textit{{Physarum} polycephalum} : {Smart} {Network} {Adaptation}},\ }\href@noop {} {\bibfield  {journal} {\bibinfo  {journal} {Annual Review of Condensed Matter Physics}\ }\textbf {\bibinfo {volume} {15}},\ \bibinfo {pages} {263} (\bibinfo {year} {2024})}\BibitemShut {NoStop}%
\bibitem [{\citenamefont {Oyarte~Galvez}\ \emph {et~al.}(2025)\citenamefont {Oyarte~Galvez}, \citenamefont {Bisot}, \citenamefont {Bourrianne}, \citenamefont {Cargill}, \citenamefont {Klein}, \citenamefont {Van~Son}, \citenamefont {Van~Krugten}, \citenamefont {Caldas}, \citenamefont {Clerc}, \citenamefont {Lin}, \citenamefont {Kahane}, \citenamefont {Van~Staalduine}, \citenamefont {Stewart}, \citenamefont {Terry}, \citenamefont {Turcu}, \citenamefont {Van~Otterdijk}, \citenamefont {Babu}, \citenamefont {Kamp}, \citenamefont {Seynen}, \citenamefont {Steenbeek}, \citenamefont {Zomerdijk}, \citenamefont {Tutucci}, \citenamefont {Sheldrake}, \citenamefont {Godin}, \citenamefont {Kokkoris}, \citenamefont {Stone}, \citenamefont {Kiers},\ and\ \citenamefont {Shimizu}}]{oyarte_galvez_travelling-wave_2025}%
  \BibitemOpen
  \bibfield  {author} {\bibinfo {author} {\bibfnamefont {L.}~\bibnamefont {Oyarte~Galvez}}, \bibinfo {author} {\bibfnamefont {C.}~\bibnamefont {Bisot}}, \bibinfo {author} {\bibfnamefont {P.}~\bibnamefont {Bourrianne}}, \bibinfo {author} {\bibfnamefont {R.}~\bibnamefont {Cargill}}, \bibinfo {author} {\bibfnamefont {M.}~\bibnamefont {Klein}}, \bibinfo {author} {\bibfnamefont {M.}~\bibnamefont {Van~Son}}, \bibinfo {author} {\bibfnamefont {J.}~\bibnamefont {Van~Krugten}}, \bibinfo {author} {\bibfnamefont {V.}~\bibnamefont {Caldas}}, \bibinfo {author} {\bibfnamefont {T.}~\bibnamefont {Clerc}}, \bibinfo {author} {\bibfnamefont {K.-K.}\ \bibnamefont {Lin}}, \bibinfo {author} {\bibfnamefont {F.}~\bibnamefont {Kahane}}, \bibinfo {author} {\bibfnamefont {S.}~\bibnamefont {Van~Staalduine}}, \bibinfo {author} {\bibfnamefont {J.~D.}\ \bibnamefont {Stewart}}, \bibinfo {author} {\bibfnamefont {V.}~\bibnamefont {Terry}}, \bibinfo {author} {\bibfnamefont {B.}~\bibnamefont {Turcu}}, \bibinfo {author} {\bibfnamefont
  {S.}~\bibnamefont {Van~Otterdijk}}, \bibinfo {author} {\bibfnamefont {A.}~\bibnamefont {Babu}}, \bibinfo {author} {\bibfnamefont {M.}~\bibnamefont {Kamp}}, \bibinfo {author} {\bibfnamefont {M.}~\bibnamefont {Seynen}}, \bibinfo {author} {\bibfnamefont {B.}~\bibnamefont {Steenbeek}}, \bibinfo {author} {\bibfnamefont {J.}~\bibnamefont {Zomerdijk}}, \bibinfo {author} {\bibfnamefont {E.}~\bibnamefont {Tutucci}}, \bibinfo {author} {\bibfnamefont {M.}~\bibnamefont {Sheldrake}}, \bibinfo {author} {\bibfnamefont {C.}~\bibnamefont {Godin}}, \bibinfo {author} {\bibfnamefont {V.}~\bibnamefont {Kokkoris}}, \bibinfo {author} {\bibfnamefont {H.~A.}\ \bibnamefont {Stone}}, \bibinfo {author} {\bibfnamefont {E.~T.}\ \bibnamefont {Kiers}},\ and\ \bibinfo {author} {\bibfnamefont {T.~S.}\ \bibnamefont {Shimizu}},\ }\bibfield  {title} {\bibinfo {title} {A travelling-wave strategy for plant–fungal trade},\ }\href@noop {} {\bibfield  {journal} {\bibinfo  {journal} {Nature}\ }\textbf {\bibinfo {volume} {639}},\ \bibinfo {pages}
  {172} (\bibinfo {year} {2025})}\BibitemShut {NoStop}%
\bibitem [{\citenamefont {Ogawa}\ \emph {et~al.}(2025)\citenamefont {Ogawa}, \citenamefont {Koyama}, \citenamefont {Omori}, \citenamefont {Kikuchi}, \citenamefont {de~Maleprade}, \citenamefont {Goldstein},\ and\ \citenamefont {Ishikawa}}]{ogawa2025architecture}%
  \BibitemOpen
  \bibfield  {author} {\bibinfo {author} {\bibfnamefont {T.}\ \bibnamefont {Ogawa}}, \bibinfo {author} {\bibfnamefont {S.}\ \bibnamefont {Koyama}}, \bibinfo {author} {\bibfnamefont {T.}\ \bibnamefont {Omori}}, \bibinfo {author} {\bibfnamefont {K.}\ \bibnamefont {Kikuchi}}, \bibinfo {author} {\bibfnamefont {H.}\ \bibnamefont {de~Maleprade}}, \bibinfo {author} {\bibfnamefont {R.~E.}\ \bibnamefont {Goldstein}},\ and\ \bibinfo {author} {\bibfnamefont {T.}\ \bibnamefont {Ishikawa}},}%
  \bibfield  {title} {\bibinfo {title} {The architecture of sponge choanocyte chambers is well adapted to mechanical pumping functions},}%
  \href {https://doi.org/10.1073/pnas.2421296122} {\bibfield  {journal} {\bibinfo  {journal} { Proceedings of the National Academy of Sciences}\ }\textbf {\bibinfo {volume} {122}},\ \bibinfo {pages} {e2421296122} (\bibinfo {year} {2025})}
  \BibitemShut {NoStop}%
\bibitem [{\citenamefont {Jensen}\ \emph {et~al.}(2016)\citenamefont {Jensen}, \citenamefont {Berg-Sørensen}, \citenamefont {Bruus}, \citenamefont {Holbrook}, \citenamefont {Liesche}, \citenamefont {Schulz}, \citenamefont {Zwieniecki},\ and\ \citenamefont {Bohr}}]{jensen_sap_2016}%
  \BibitemOpen
  \bibfield  {author} {\bibinfo {author} {\bibfnamefont {K.}~\bibnamefont {Jensen}}, \bibinfo {author} {\bibfnamefont {K.}~\bibnamefont {Berg-Sørensen}}, \bibinfo {author} {\bibfnamefont {H.}~\bibnamefont {Bruus}}, \bibinfo {author} {\bibfnamefont {N.}~\bibnamefont {Holbrook}}, \bibinfo {author} {\bibfnamefont {J.}~\bibnamefont {Liesche}}, \bibinfo {author} {\bibfnamefont {A.}~\bibnamefont {Schulz}}, \bibinfo {author} {\bibfnamefont {M.}~\bibnamefont {Zwieniecki}},\ and\ \bibinfo {author} {\bibfnamefont {T.}~\bibnamefont {Bohr}},\ }\bibfield  {title} {\bibinfo {title} {Sap flow and sugar transport in plants},\ }\href {https://doi.org/10.1103/RevModPhys.88.035007} {\bibfield  {journal} {\bibinfo  {journal} {Rev. Mod. Phys.}\ }\textbf {\bibinfo {volume} {88}},\ \bibinfo {pages} {035007} (\bibinfo {year} {2016})}\BibitemShut {NoStop}%
\bibitem [{\citenamefont {Alim}\ \emph {et~al.}(2017)\citenamefont {Alim}, \citenamefont {Andrew}, \citenamefont {Pringle},\ and\ \citenamefont {Brenner}}]{alim_mechanism_2017}%
  \BibitemOpen
  \bibfield  {author} {\bibinfo {author} {\bibfnamefont {K.}~\bibnamefont {Alim}}, \bibinfo {author} {\bibfnamefont {N.}~\bibnamefont {Andrew}}, \bibinfo {author} {\bibfnamefont {A.}~\bibnamefont {Pringle}},\ and\ \bibinfo {author} {\bibfnamefont {M.~P.}\ \bibnamefont {Brenner}},\ }\bibfield  {title} {\bibinfo {title} {Mechanism of signal propagation in \textit{{Physarum} polycephalum}},\ }\href {https://doi.org/10.1073/pnas.1618114114} {\bibfield  {journal} {\bibinfo  {journal} {Proceedings of the National Academy of Sciences}\ }\textbf {\bibinfo {volume} {114}},\ \bibinfo {pages} {5136} (\bibinfo {year} {2017})}\BibitemShut {NoStop}%
\bibitem [{\citenamefont {Liese}\ \emph {et~al.}(2021)\citenamefont {Liese}, \citenamefont {Mahadevan},\ and\ \citenamefont {Carlson}}]{liese_balancing_2021}%
  \BibitemOpen
  \bibfield  {author} {\bibinfo {author} {\bibfnamefont {S.}~\bibnamefont {Liese}}, \bibinfo {author} {\bibfnamefont {L.}~\bibnamefont {Mahadevan}},\ and\ \bibinfo {author} {\bibfnamefont {A.}~\bibnamefont {Carlson}},\ }\bibfield  {title} {\bibinfo {title} {Balancing efficiency and homogeneity of biomaterial transport in networks},\ }\href {https://doi.org/10.1209/0295-5075/135/58001} {\bibfield  {journal} {\bibinfo  {journal} {EPL}\ }\textbf {\bibinfo {volume} {135}},\ \bibinfo {pages} {58001} (\bibinfo {year} {2021})}\BibitemShut {NoStop}%
\bibitem [{\citenamefont {Pereira}\ \emph {et~al.}(2023)\citenamefont {Pereira}, \citenamefont {Kaack}, \citenamefont {Guan}, \citenamefont {Silva}, \citenamefont {Miranda}, \citenamefont {Pires}, \citenamefont {Ribeiro}, \citenamefont {Schenk},\ and\ \citenamefont {Jansen}}]{pereira_angiosperms_2023}%
  \BibitemOpen
  \bibfield  {author} {\bibinfo {author} {\bibfnamefont {L.}~\bibnamefont {Pereira}}, \bibinfo {author} {\bibfnamefont {L.}~\bibnamefont {Kaack}}, \bibinfo {author} {\bibfnamefont {X.}~\bibnamefont {Guan}}, \bibinfo {author} {\bibfnamefont {L.~D.~M.}\ \bibnamefont {Silva}}, \bibinfo {author} {\bibfnamefont {M.~T.}\ \bibnamefont {Miranda}}, \bibinfo {author} {\bibfnamefont {G.~S.}\ \bibnamefont {Pires}}, \bibinfo {author} {\bibfnamefont {R.~V.}\ \bibnamefont {Ribeiro}}, \bibinfo {author} {\bibfnamefont {H.~J.}\ \bibnamefont {Schenk}},\ and\ \bibinfo {author} {\bibfnamefont {S.}~\bibnamefont {Jansen}},\ }\bibfield  {title} {\bibinfo {title} {Angiosperms follow a convex trade‐off to optimize hydraulic safety and efficiency},\ }\href {https://doi.org/10.1111/nph.19253} {\bibfield  {journal} {\bibinfo  {journal} {New Phytologist}\ }\textbf {\bibinfo {volume} {240}},\ \bibinfo {pages} {1788} (\bibinfo {year} {2023})}\BibitemShut {NoStop}%
\bibitem [{\citenamefont {Heil}\ and\ \citenamefont {Hazel}(2011)}]{heil_fluid-structure_2011}%
  \BibitemOpen
  \bibfield  {author} {\bibinfo {author} {\bibfnamefont {M.}~\bibnamefont {Heil}}\ and\ \bibinfo {author} {\bibfnamefont {A.~L.}\ \bibnamefont {Hazel}},\ }\bibfield  {title} {\bibinfo {title} {Fluid-structure interaction in internal physiological flows},\ }\href@noop {} {\bibfield  {journal} {\bibinfo  {journal} {Annu. Rev. Fluid Mech.}\ }\textbf {\bibinfo {volume} {43}},\ \bibinfo {pages} {141} (\bibinfo {year} {2011})}\BibitemShut {NoStop}%
\bibitem [{\citenamefont {Fallahi}\ \emph {et~al.}(2019)\citenamefont {Fallahi}, \citenamefont {Zhang}, \citenamefont {Phan},\ and\ \citenamefont {Nguyen}}]{fallahi2019flexible}%
  \BibitemOpen
  \bibfield  {author} {\bibinfo {author} {\bibfnamefont {H.}~\bibnamefont {Fallahi}}, \bibinfo {author} {\bibfnamefont {J.}~\bibnamefont {Zhang}}, \bibinfo {author} {\bibfnamefont {H.-P.}\ \bibnamefont {Phan}},\ and\ \bibinfo {author} {\bibfnamefont {N.-T.}\ \bibnamefont {Nguyen}},\ }\bibfield  {title} {\bibinfo {title} {Flexible microfluidics: Fundamentals, recent developments, and applications},\ }\href@noop {} {\bibfield  {journal} {\bibinfo  {journal} {Micromachines}\ }\textbf {\bibinfo {volume} {10}},\ \bibinfo {pages} {830} (\bibinfo {year} {2019})}\BibitemShut {NoStop}%
\bibitem [{\citenamefont {Battat}\ \emph {et~al.}(2022)\citenamefont {Battat}, \citenamefont {Weitz},\ and\ \citenamefont {Whitesides}}]{battat_nonlinear_2022}%
  \BibitemOpen
  \bibfield  {author} {\bibinfo {author} {\bibfnamefont {S.}~\bibnamefont {Battat}}, \bibinfo {author} {\bibfnamefont {D.~A.}\ \bibnamefont {Weitz}},\ and\ \bibinfo {author} {\bibfnamefont {G.~M.}\ \bibnamefont {Whitesides}},\ }\bibfield  {title} {\bibinfo {title} {Nonlinear {Phenomena} in {Microfluidics}},\ }\href {https://doi.org/10.1021/acs.chemrev.1c00985} {\bibfield  {journal} {\bibinfo  {journal} {Chemical Reviews}\ }\textbf {\bibinfo {volume} {122}},\ \bibinfo {pages} {6921} (\bibinfo {year} {2022})}\BibitemShut {NoStop}%
\bibitem [{\citenamefont {Xia}\ \emph {et~al.}(2021)\citenamefont {Xia}, \citenamefont {Wu}, \citenamefont {Zheng}, \citenamefont {Zhang},\ and\ \citenamefont {Wang}}]{xia_nonlinear_2021}%
  \BibitemOpen
  \bibfield  {author} {\bibinfo {author} {\bibfnamefont {H.~M.}\ \bibnamefont {Xia}}, \bibinfo {author} {\bibfnamefont {J.~W.}\ \bibnamefont {Wu}}, \bibinfo {author} {\bibfnamefont {J.~J.}\ \bibnamefont {Zheng}}, \bibinfo {author} {\bibfnamefont {J.}~\bibnamefont {Zhang}},\ and\ \bibinfo {author} {\bibfnamefont {Z.~P.}\ \bibnamefont {Wang}},\ }\bibfield  {title} {\bibinfo {title} {Nonlinear microfluidics: device physics, functions, and applications},\ }\href {https://doi.org/10.1039/D0LC01120G} {\bibfield  {journal} {\bibinfo  {journal} {Lab on a Chip}\ }\textbf {\bibinfo {volume} {21}},\ \bibinfo {pages} {1241} (\bibinfo {year} {2021})}\BibitemShut {NoStop}%
\bibitem [{\citenamefont {Case}\ \emph {et~al.}(2020)\citenamefont {Case}, \citenamefont {Angilella},\ and\ \citenamefont {Motter}}]{case_spontaneous_2020}%
  \BibitemOpen
  \bibfield  {author} {\bibinfo {author} {\bibfnamefont {D.~J.}\ \bibnamefont {Case}}, \bibinfo {author} {\bibfnamefont {J.-R.}\ \bibnamefont {Angilella}},\ and\ \bibinfo {author} {\bibfnamefont {A.~E.}\ \bibnamefont {Motter}},\ }\bibfield  {title} {\bibinfo {title} {Spontaneous oscillations and negative-conductance transitions in microfluidic networks},\ }\href {https://doi.org/10.1126/sciadv.aay6761} {\bibfield  {journal} {\bibinfo  {journal} {Science Advances}\ }\textbf {\bibinfo {volume} {6}},\ \bibinfo {pages} {eaay6761} (\bibinfo {year} {2020})}\BibitemShut {NoStop}%
\bibitem [{\citenamefont {Brandenbourger}\ \emph {et~al.}(2020)\citenamefont {Brandenbourger}, \citenamefont {Dangremont}, \citenamefont {Sprik},\ and\ \citenamefont {Coulais}}]{brandenbourger_tunable_2020}%
  \BibitemOpen
  \bibfield  {author} {\bibinfo {author} {\bibfnamefont {M.}~\bibnamefont {Brandenbourger}}, \bibinfo {author} {\bibfnamefont {A.}~\bibnamefont {Dangremont}}, \bibinfo {author} {\bibfnamefont {R.}~\bibnamefont {Sprik}},\ and\ \bibinfo {author} {\bibfnamefont {C.}~\bibnamefont {Coulais}},\ }\bibfield  {title} {\bibinfo {title} {Tunable flow asymmetry and flow rectification with bio-inspired soft leaflets},\ }\href {https://doi.org/10.1103/PhysRevFluids.5.084102} {\bibfield  {journal} {\bibinfo  {journal} {Phys. Rev. Fluids}\ }\textbf {\bibinfo {volume} {5}},\ \bibinfo {pages} {084102} (\bibinfo {year} {2020})}\BibitemShut {NoStop}%
\bibitem [{\citenamefont {Park}\ \emph {et~al.}(2021)\citenamefont {Park}, \citenamefont {Tixier}, \citenamefont {Paludan}, \citenamefont {Østergaard}, \citenamefont {Zwieniecki},\ and\ \citenamefont {Jensen}}]{park_fluid-structure_2021}%
  \BibitemOpen
  \bibfield  {author} {\bibinfo {author} {\bibfnamefont {K.}~\bibnamefont {Park}}, \bibinfo {author} {\bibfnamefont {A.}~\bibnamefont {Tixier}}, \bibinfo {author} {\bibfnamefont {M.}~\bibnamefont {Paludan}}, \bibinfo {author} {\bibfnamefont {E.}~\bibnamefont {Østergaard}}, \bibinfo {author} {\bibfnamefont {M.}~\bibnamefont {Zwieniecki}},\ and\ \bibinfo {author} {\bibfnamefont {K.~H.}\ \bibnamefont {Jensen}},\ }\bibfield  {title} {\bibinfo {title} {Fluid-structure interactions enable passive flow control in real and biomimetic plants},\ }\href {https://doi.org/10.1103/PhysRevFluids.6.123102} {\bibfield  {journal} {\bibinfo  {journal} {Phys. Rev. Fluids}\ }\textbf {\bibinfo {volume} {6}},\ \bibinfo {pages} {123102} (\bibinfo {year} {2021})}\BibitemShut {NoStop}%
\bibitem [{\citenamefont {Jambon-Puillet}(2025)}]{jambonpuillet2025densearrayelastichairs}%
  \BibitemOpen
  \bibfield  {author} {\bibinfo {author} {\bibfnamefont {E.}~\bibnamefont {Jambon-Puillet}},\ }\bibfield  {title} {\bibinfo {title} {Dense array of elastic hairs obstructing a fluidic channel},\ }\href {https://arxiv.org/abs/2501.01875} {\bibfield  {journal} {\bibinfo  {journal} {arXiv}\ } (\bibinfo {year} {2025})},\ \Eprint {https://arxiv.org/abs/2501.01875} {arXiv:2501.01875 [cond-mat.soft]} \BibitemShut {NoStop}%
\bibitem [{\citenamefont {Case}\ \emph {et~al.}(2019)\citenamefont {Case}, \citenamefont {Liu}, \citenamefont {Kiss}, \citenamefont {Angilella},\ and\ \citenamefont {Motter}}]{case_braesss_2019}%
  \BibitemOpen
  \bibfield  {author} {\bibinfo {author} {\bibfnamefont {D.~J.}\ \bibnamefont {Case}}, \bibinfo {author} {\bibfnamefont {Y.}~\bibnamefont {Liu}}, \bibinfo {author} {\bibfnamefont {I.~Z.}\ \bibnamefont {Kiss}}, \bibinfo {author} {\bibfnamefont {J.-R.}\ \bibnamefont {Angilella}},\ and\ \bibinfo {author} {\bibfnamefont {A.~E.}\ \bibnamefont {Motter}},\ }\bibfield  {title} {\bibinfo {title} {Braess’s paradox and programmable behaviour in microfluidic networks},\ }\href {https://doi.org/10.1038/s41586-019-1701-6} {\bibfield  {journal} {\bibinfo  {journal} {Nature}\ }\textbf {\bibinfo {volume} {574}},\ \bibinfo {pages} {647} (\bibinfo {year} {2019})}\BibitemShut {NoStop}%
\bibitem [{\citenamefont {Martínez-Calvo}\ \emph {et~al.}(2024)\citenamefont {Martínez-Calvo}, \citenamefont {Biviano}, \citenamefont {Christensen}, \citenamefont {Katifori}, \citenamefont {Jensen},\ and\ \citenamefont {Ruiz-García}}]{martinez-calvo_fluidic_2024}%
  \BibitemOpen
  \bibfield  {author} {\bibinfo {author} {\bibfnamefont {A.}~\bibnamefont {Martínez-Calvo}}, \bibinfo {author} {\bibfnamefont {M.~D.}\ \bibnamefont {Biviano}}, \bibinfo {author} {\bibfnamefont {A.~H.}\ \bibnamefont {Christensen}}, \bibinfo {author} {\bibfnamefont {E.}~\bibnamefont {Katifori}}, \bibinfo {author} {\bibfnamefont {K.~H.}\ \bibnamefont {Jensen}},\ and\ \bibinfo {author} {\bibfnamefont {M.}~\bibnamefont {Ruiz-García}},\ }\bibfield  {title} {\bibinfo {title} {The fluidic memristor as a collective phenomenon in elastohydrodynamic networks},\ }\href {https://doi.org/10.1038/s41467-024-47110-0} {\bibfield  {journal} {\bibinfo  {journal} {Nat Commun}\ }\textbf {\bibinfo {volume} {15}},\ \bibinfo {pages} {3121} (\bibinfo {year} {2024})}\BibitemShut {NoStop}%
\bibitem [{\citenamefont {Bradley}\ \emph {et~al.}(2019)\citenamefont {Bradley}, \citenamefont {Box}, \citenamefont {Hewitt},\ and\ \citenamefont {Vella}}]{bradley_wettability-independent_2019}%
  \BibitemOpen
  \bibfield  {author} {\bibinfo {author} {\bibfnamefont {A.~T.}\ \bibnamefont {Bradley}}, \bibinfo {author} {\bibfnamefont {F.}~\bibnamefont {Box}}, \bibinfo {author} {\bibfnamefont {I.~J.}\ \bibnamefont {Hewitt}},\ and\ \bibinfo {author} {\bibfnamefont {D.}~\bibnamefont {Vella}},\ }\bibfield  {title} {\bibinfo {title} {Wettability-{Independent} {Droplet} {Transport} by \textit{{Bendotaxis}}},\ }\href {https://doi.org/10.1103/PhysRevLett.122.074503} {\bibfield  {journal} {\bibinfo  {journal} {Physical Review Letters}\ }\textbf {\bibinfo {volume} {122}},\ \bibinfo {pages} {074503} (\bibinfo {year} {2019})}\BibitemShut {NoStop}%
\bibitem [{\citenamefont {Bradley}\ \emph {et~al.}(2023)\citenamefont {Bradley}, \citenamefont {Hewitt},\ and\ \citenamefont {Vella}}]{bradley_bendocapillary_2023}%
  \BibitemOpen
  \bibfield  {author} {\bibinfo {author} {\bibfnamefont {A.~T.}\ \bibnamefont {Bradley}}, \bibinfo {author} {\bibfnamefont {I.~J.}\ \bibnamefont {Hewitt}},\ and\ \bibinfo {author} {\bibfnamefont {D.}~\bibnamefont {Vella}},\ }\bibfield  {title} {\bibinfo {title} {Bendocapillary instability of liquid in a flexible-walled channel},\ }\href {https://doi.org/10.1017/jfm.2022.1025} {\bibfield  {journal} {\bibinfo  {journal} {Journal of Fluid Mechanics}\ }\textbf {\bibinfo {volume} {955}},\ \bibinfo {pages} {A26} (\bibinfo {year} {2023})}\BibitemShut {NoStop}%
\bibitem [{\citenamefont {Ushay}\ \emph {et~al.}(2023)\citenamefont {Ushay}, \citenamefont {Jambon-Puillet},\ and\ \citenamefont {Brun}}]{ushay2023interfacial}%
  \BibitemOpen
  \bibfield  {author} {\bibinfo {author} {\bibfnamefont {C.}~\bibnamefont {Ushay}}, \bibinfo {author} {\bibfnamefont {E.}~\bibnamefont {Jambon-Puillet}},\ and\ \bibinfo {author} {\bibfnamefont {P.-T.}\ \bibnamefont {Brun}},\ }\bibfield  {title} {\bibinfo {title} {Interfacial flows past arrays of elastic fibers},\ }\href@noop {} {\bibfield  {journal} {\bibinfo  {journal} {Physical Review Fluids}\ }\textbf {\bibinfo {volume} {8}},\ \bibinfo {pages} {044001} (\bibinfo {year} {2023})}\BibitemShut {NoStop}%
\bibitem [{\citenamefont {Pihler-Puzović}\ \emph {et~al.}(2012)\citenamefont {Pihler-Puzović}, \citenamefont {Illien}, \citenamefont {Heil},\ and\ \citenamefont {Juel}}]{pihler-puzovic_suppression_2012}%
  \BibitemOpen
  \bibfield  {author} {\bibinfo {author} {\bibfnamefont {D.}~\bibnamefont {Pihler-Puzović}}, \bibinfo {author} {\bibfnamefont {P.}~\bibnamefont {Illien}}, \bibinfo {author} {\bibfnamefont {M.}~\bibnamefont {Heil}},\ and\ \bibinfo {author} {\bibfnamefont {A.}~\bibnamefont {Juel}},\ }\bibfield  {title} {\bibinfo {title} {Suppression of complex fingerlike patterns at the interface between air and a viscous fluid by elastic membranes},\ }\href@noop {} {\bibfield  {journal} {\bibinfo  {journal} {Phys. Rev. Lett.}\ }\textbf {\bibinfo {volume} {108}},\ \bibinfo {pages} {074502} (\bibinfo {year} {2012})}\BibitemShut {NoStop}%
\bibitem [{\citenamefont {Juel}\ \emph {et~al.}(2018)\citenamefont {Juel}, \citenamefont {Pihler-Puzović},\ and\ \citenamefont {Heil}}]{juel_instabilities_2018}%
  \BibitemOpen
  \bibfield  {author} {\bibinfo {author} {\bibfnamefont {A.}~\bibnamefont {Juel}}, \bibinfo {author} {\bibfnamefont {D.}~\bibnamefont {Pihler-Puzović}},\ and\ \bibinfo {author} {\bibfnamefont {M.}~\bibnamefont {Heil}},\ }\bibfield  {title} {\bibinfo {title} {Instabilities in blistering},\ }\href@noop {} {\bibfield  {journal} {\bibinfo  {journal} {Annu. Rev. Fluid Mech.}\ }\textbf {\bibinfo {volume} {50}},\ \bibinfo {pages} {691} (\bibinfo {year} {2018})}\BibitemShut {NoStop}%
\bibitem [{\citenamefont {Louf}\ \emph {et~al.}(2020)\citenamefont {Louf}, \citenamefont {Knoblauch},\ and\ \citenamefont {Jensen}}]{louf_bending_2020}%
  \BibitemOpen
  \bibfield  {author} {\bibinfo {author} {\bibfnamefont {J.-F.}\ \bibnamefont {Louf}}, \bibinfo {author} {\bibfnamefont {J.}~\bibnamefont {Knoblauch}},\ and\ \bibinfo {author} {\bibfnamefont {K.~H.}\ \bibnamefont {Jensen}},\ }\bibfield  {title} {\bibinfo {title} {Bending and {Stretching} of {Soft} {Pores} {Enable} {Passive} {Control} of {Fluid} {Flows}},\ }\href {https://doi.org/10.1103/PhysRevLett.125.098101} {\bibfield  {journal} {\bibinfo  {journal} {Physical Review Letters}\ }\textbf {\bibinfo {volume} {125}},\ \bibinfo {pages} {098101} (\bibinfo {year} {2020})}\BibitemShut {NoStop}%
\bibitem [{\citenamefont {Cappello}\ \emph {et~al.}(2022)\citenamefont {Cappello}, \citenamefont {Du~Roure}, \citenamefont {Gallaire}, \citenamefont {Duprat},\ and\ \citenamefont {Lindner}}]{cappello_fiber_2022}%
  \BibitemOpen
  \bibfield  {author} {\bibinfo {author} {\bibfnamefont {J.}~\bibnamefont {Cappello}}, \bibinfo {author} {\bibfnamefont {O.}~\bibnamefont {Du~Roure}}, \bibinfo {author} {\bibfnamefont {F.}~\bibnamefont {Gallaire}}, \bibinfo {author} {\bibfnamefont {C.}~\bibnamefont {Duprat}},\ and\ \bibinfo {author} {\bibfnamefont {A.}~\bibnamefont {Lindner}},\ }\bibfield  {title} {\bibinfo {title} {Fiber {Buckling} in {Confined} {Viscous} {Flows}: {An} {Absolute} {Instability} {Described} by the {Linear} {Ginzburg}-{Landau} {Equation}},\ }\href {https://doi.org/10.1103/PhysRevLett.129.074504} {\bibfield  {journal} {\bibinfo  {journal} {Physical Review Letters}\ }\textbf {\bibinfo {volume} {129}},\ \bibinfo {pages} {074504} (\bibinfo {year} {2022})}\BibitemShut {NoStop}%
\bibitem [{\citenamefont {Guyard}\ \emph {et~al.}(2022)\citenamefont {Guyard}, \citenamefont {Restagno},\ and\ \citenamefont {McGraw}}]{guyard2022elastohydrodynamic}%
  \BibitemOpen
  \bibfield  {author} {\bibinfo {author} {\bibfnamefont {G.}~\bibnamefont {Guyard}}, \bibinfo {author} {\bibfnamefont {F.}~\bibnamefont {Restagno}},\ and\ \bibinfo {author} {\bibfnamefont {J.~D.}\ \bibnamefont {McGraw}},\ }\bibfield  {title} {\bibinfo {title} {Elastohydrodynamic relaxation of soft and deformable microchannels},\ }\href@noop {} {\bibfield  {journal} {\bibinfo  {journal} {Phys. Rev. Lett.}\ }\textbf {\bibinfo {volume} {129}},\ \bibinfo {pages} {204501} (\bibinfo {year} {2022})}\BibitemShut {NoStop}%
\bibitem [{\citenamefont {Paludan}\ \emph {et~al.}(2024)\citenamefont {Paludan}, \citenamefont {Dollet}, \citenamefont {Marmottant},\ and\ \citenamefont {Jensen}}]{paludan_elastohydrodynamic_2024}%
  \BibitemOpen
  \bibfield  {author} {\bibinfo {author} {\bibfnamefont {M.~V.}\ \bibnamefont {Paludan}}, \bibinfo {author} {\bibfnamefont {B.}~\bibnamefont {Dollet}}, \bibinfo {author} {\bibfnamefont {P.}~\bibnamefont {Marmottant}},\ and\ \bibinfo {author} {\bibfnamefont {K.~H.}\ \bibnamefont {Jensen}},\ }\bibfield  {title} {\bibinfo {title} {Elastohydrodynamic interactions in soft hydraulic knots},\ }\href {https://doi.org/10.1017/jfm.2024.239} {\bibfield  {journal} {\bibinfo  {journal} {J. Fluid Mech.}\ }\textbf {\bibinfo {volume} {984}},\ \bibinfo {pages} {A55} (\bibinfo {year} {2024})}\BibitemShut {NoStop}%
\bibitem [{\citenamefont {Garg}\ \emph {et~al.}(2024)\citenamefont {Garg}, \citenamefont {Ledda}, \citenamefont {Pedersen},\ and\ \citenamefont {Pezzulla}}]{garg_passive_2024}%
  \BibitemOpen
  \bibfield  {author} {\bibinfo {author} {\bibfnamefont {H.}~\bibnamefont {Garg}}, \bibinfo {author} {\bibfnamefont {P.~G.}\ \bibnamefont {Ledda}}, \bibinfo {author} {\bibfnamefont {J.~S.}\ \bibnamefont {Pedersen}},\ and\ \bibinfo {author} {\bibfnamefont {M.}~\bibnamefont {Pezzulla}},\ }\bibfield  {title} {\bibinfo {title} {Passive {Viscous} {Flow} {Selection} via {Fluid}-{Induced} {Buckling}},\ }\href {https://doi.org/10.1103/PhysRevLett.133.084001} {\bibfield  {journal} {\bibinfo  {journal} {Physical Review Letters}\ }\textbf {\bibinfo {volume} {133}},\ \bibinfo {pages} {084001} (\bibinfo {year} {2024})}\BibitemShut {NoStop}%
\bibitem [{\citenamefont {Tyree}\ and\ \citenamefont {Zimmermann}(2013)}]{tyree_xylem_2013}%
  \BibitemOpen
  \bibfield  {author} {\bibinfo {author} {\bibfnamefont {M.~T.}\ \bibnamefont {Tyree}}\ and\ \bibinfo {author} {\bibfnamefont {M.~H.}\ \bibnamefont {Zimmermann}},\ }\href@noop {} {\emph {\bibinfo {title} {Xylem {Structure} and the {Ascent} of {Sap}}}}\ (\bibinfo  {publisher} {Springer Science \& Business Media},\ \bibinfo {year} {2013})\BibitemShut {NoStop}%
\bibitem [{\citenamefont {Stroock}\ \emph {et~al.}(2014)\citenamefont {Stroock}, \citenamefont {Pagay}, \citenamefont {Zwieniecki},\ and\ \citenamefont {Michele~Holbrook}}]{stroock_physicochemical_2014}%
  \BibitemOpen
  \bibfield  {author} {\bibinfo {author} {\bibfnamefont {A.~D.}\ \bibnamefont {Stroock}}, \bibinfo {author} {\bibfnamefont {V.~V.}\ \bibnamefont {Pagay}}, \bibinfo {author} {\bibfnamefont {M.~A.}\ \bibnamefont {Zwieniecki}},\ and\ \bibinfo {author} {\bibfnamefont {N.}~\bibnamefont {Michele~Holbrook}},\ }\bibfield  {title} {\bibinfo {title} {The {Physicochemical} {Hydrodynamics} of {Vascular} {Plants}},\ }\href {https://doi.org/10.1146/annurev-fluid-010313-141411} {\bibfield  {journal} {\bibinfo  {journal} {Annu. Rev. Fluid Mech.}\ }\textbf {\bibinfo {volume} {46}},\ \bibinfo {pages} {615} (\bibinfo {year} {2014})}\BibitemShut {NoStop}%
\bibitem [{\citenamefont {Katifori}(2018)}]{katifori_transport_2018}%
  \BibitemOpen
  \bibfield  {author} {\bibinfo {author} {\bibfnamefont {E.}~\bibnamefont {Katifori}},\ }\bibfield  {title} {\bibinfo {title} {The transport network of a leaf},\ }\href {https://doi.org/10.1016/j.crhy.2018.10.007} {\bibfield  {journal} {\bibinfo  {journal} {Comptes Rendus Physique}\ }\textbf {\bibinfo {volume} {19}},\ \bibinfo {pages} {244} (\bibinfo {year} {2018})}\BibitemShut {NoStop}%
\bibitem [{\citenamefont {Choat}\ \emph {et~al.}(2012)\citenamefont {Choat}, \citenamefont {Jansen}, \citenamefont {Brodribb}, \citenamefont {Cochard}, \citenamefont {Delzon}, \citenamefont {Bhaskar}, \citenamefont {Bucci}, \citenamefont {Feild}, \citenamefont {Gleason}, \citenamefont {Hacke} \emph {et~al.}}]{choat2012global}%
  \BibitemOpen
  \bibfield  {author} {\bibinfo {author} {\bibfnamefont {B.}~\bibnamefont {Choat}}, \bibinfo {author} {\bibfnamefont {S.}~\bibnamefont {Jansen}}, \bibinfo {author} {\bibfnamefont {T.~J.}\ \bibnamefont {Brodribb}}, \bibinfo {author} {\bibfnamefont {H.}~\bibnamefont {Cochard}}, \bibinfo {author} {\bibfnamefont {S.}~\bibnamefont {Delzon}}, \bibinfo {author} {\bibfnamefont {R.}~\bibnamefont {Bhaskar}}, \bibinfo {author} {\bibfnamefont {S.~J.}\ \bibnamefont {Bucci}}, \bibinfo {author} {\bibfnamefont {T.~S.}\ \bibnamefont {Feild}}, \bibinfo {author} {\bibfnamefont {S.~M.}\ \bibnamefont {Gleason}}, \bibinfo {author} {\bibfnamefont {U.~G.}\ \bibnamefont {Hacke}}, \emph {et~al.},\ }\bibfield  {title} {\bibinfo {title} {Global convergence in the vulnerability of forests to drought},\ }\href@noop {} {\bibfield  {journal} {\bibinfo  {journal} {Nature}\ }\textbf {\bibinfo {volume} {491}},\ \bibinfo {pages} {752} (\bibinfo {year} {2012})}\BibitemShut {NoStop}%
\bibitem [{\citenamefont {Choat}\ \emph {et~al.}(2018)\citenamefont {Choat}, \citenamefont {Brodribb}, \citenamefont {Brodersen}, \citenamefont {Duursma}, \citenamefont {L{\'o}pez},\ and\ \citenamefont {Medlyn}}]{choat2018triggers}%
  \BibitemOpen
  \bibfield  {author} {\bibinfo {author} {\bibfnamefont {B.}~\bibnamefont {Choat}}, \bibinfo {author} {\bibfnamefont {T.~J.}\ \bibnamefont {Brodribb}}, \bibinfo {author} {\bibfnamefont {C.~R.}\ \bibnamefont {Brodersen}}, \bibinfo {author} {\bibfnamefont {R.~A.}\ \bibnamefont {Duursma}}, \bibinfo {author} {\bibfnamefont {R.}~\bibnamefont {L{\'o}pez}},\ and\ \bibinfo {author} {\bibfnamefont {B.~E.}\ \bibnamefont {Medlyn}},\ }\bibfield  {title} {\bibinfo {title} {Triggers of tree mortality under drought},\ }\href@noop {} {\bibfield  {journal} {\bibinfo  {journal} {Nature}\ }\textbf {\bibinfo {volume} {558}},\ \bibinfo {pages} {531} (\bibinfo {year} {2018})}\BibitemShut {NoStop}%
\bibitem [{\citenamefont {Torres-Ruiz}\ \emph {et~al.}(2024)\citenamefont {Torres-Ruiz}, \citenamefont {Cochard}, \citenamefont {Delzon}, \citenamefont {Boivin}, \citenamefont {Burlett}, \citenamefont {Cailleret}, \citenamefont {Corso}, \citenamefont {Delmas}, \citenamefont {De~Caceres}, \citenamefont {Diaz-Espejo} \emph {et~al.}}]{torres2024plant}%
  \BibitemOpen
  \bibfield  {author} {\bibinfo {author} {\bibfnamefont {J.~M.}\ \bibnamefont {Torres-Ruiz}}, \bibinfo {author} {\bibfnamefont {H.}~\bibnamefont {Cochard}}, \bibinfo {author} {\bibfnamefont {S.}~\bibnamefont {Delzon}}, \bibinfo {author} {\bibfnamefont {T.}~\bibnamefont {Boivin}}, \bibinfo {author} {\bibfnamefont {R.}~\bibnamefont {Burlett}}, \bibinfo {author} {\bibfnamefont {M.}~\bibnamefont {Cailleret}}, \bibinfo {author} {\bibfnamefont {D.}~\bibnamefont {Corso}}, \bibinfo {author} {\bibfnamefont {C.~E.}\ \bibnamefont {Delmas}}, \bibinfo {author} {\bibfnamefont {M.}~\bibnamefont {De~Caceres}}, \bibinfo {author} {\bibfnamefont {A.}~\bibnamefont {Diaz-Espejo}}, \emph {et~al.},\ }\bibfield  {title} {\bibinfo {title} {Plant hydraulics at the heart of plant, crops and ecosystem functions in the face of climate change},\ }\href@noop {} {\bibfield  {journal} {\bibinfo  {journal} {New Phytologist}\ }\textbf {\bibinfo {volume} {241}},\ \bibinfo {pages} {984} (\bibinfo {year} {2024})}\BibitemShut {NoStop}%
\bibitem [{\citenamefont {Brodribb}\ \emph {et~al.}(2016)\citenamefont {Brodribb}, \citenamefont {Bienaimé},\ and\ \citenamefont {Marmottant}}]{brodribb_revealing_2016}%
  \BibitemOpen
  \bibfield  {author} {\bibinfo {author} {\bibfnamefont {T.~J.}\ \bibnamefont {Brodribb}}, \bibinfo {author} {\bibfnamefont {D.}~\bibnamefont {Bienaimé}},\ and\ \bibinfo {author} {\bibfnamefont {P.}~\bibnamefont {Marmottant}},\ }\bibfield  {title} {\bibinfo {title} {Revealing catastrophic failure of leaf networks under stress},\ }\href@noop {} {\bibfield  {journal} {\bibinfo  {journal} {Proc. Natl. Acad. Sci.}\ }\textbf {\bibinfo {volume} {113}},\ \bibinfo {pages} {4865} (\bibinfo {year} {2016})}\BibitemShut {NoStop}%
\bibitem [{\citenamefont {Brodribb}\ \emph {et~al.}(2017)\citenamefont {Brodribb}, \citenamefont {Carriqui}, \citenamefont {Delzon},\ and\ \citenamefont {Lucani}}]{brodribb_optical_2017}%
  \BibitemOpen
  \bibfield  {author} {\bibinfo {author} {\bibfnamefont {T.~J.}\ \bibnamefont {Brodribb}}, \bibinfo {author} {\bibfnamefont {M.}~\bibnamefont {Carriqui}}, \bibinfo {author} {\bibfnamefont {S.}~\bibnamefont {Delzon}},\ and\ \bibinfo {author} {\bibfnamefont {C.}~\bibnamefont {Lucani}},\ }\bibfield  {title} {\bibinfo {title} {Optical measurement of stem xylem vulnerability},\ }\href@noop {} {\bibfield  {journal} {\bibinfo  {journal} {Plant Physiology}\ }\textbf {\bibinfo {volume} {174}},\ \bibinfo {pages} {2054} (\bibinfo {year} {2017})}\BibitemShut {NoStop}%
\bibitem [{\citenamefont {Cochard}\ \emph {et~al.}(2004)\citenamefont {Cochard}, \citenamefont {Froux}, \citenamefont {Mayr},\ and\ \citenamefont {Coutand}}]{cochard_xylem_2004}%
  \BibitemOpen
  \bibfield  {author} {\bibinfo {author} {\bibfnamefont {H.}~\bibnamefont {Cochard}}, \bibinfo {author} {\bibfnamefont {F.}~\bibnamefont {Froux}}, \bibinfo {author} {\bibfnamefont {S.}~\bibnamefont {Mayr}},\ and\ \bibinfo {author} {\bibfnamefont {C.}~\bibnamefont {Coutand}},\ }\bibfield  {title} {\bibinfo {title} {Xylem wall collapse in water-stressed pine needles},\ }\href@noop {} {\bibfield  {journal} {\bibinfo  {journal} {Plant Physiol.}\ }\textbf {\bibinfo {volume} {134}},\ \bibinfo {pages} {401} (\bibinfo {year} {2004})}\BibitemShut {NoStop}%
\bibitem [{\citenamefont {Zhang}\ \emph {et~al.}(2016)\citenamefont {Zhang}, \citenamefont {Rockwell}, \citenamefont {Graham}, \citenamefont {Alexander},\ and\ \citenamefont {Holbrook}}]{zhang_reversible_2016}%
  \BibitemOpen
  \bibfield  {author} {\bibinfo {author} {\bibfnamefont {Y.~J.}\ \bibnamefont {Zhang}}, \bibinfo {author} {\bibfnamefont {F.~E.}\ \bibnamefont {Rockwell}}, \bibinfo {author} {\bibfnamefont {A.~C.}\ \bibnamefont {Graham}}, \bibinfo {author} {\bibfnamefont {T.}~\bibnamefont {Alexander}},\ and\ \bibinfo {author} {\bibfnamefont {N.~M.}\ \bibnamefont {Holbrook}},\ }\bibfield  {title} {\bibinfo {title} {Reversible leaf xylem collapse: a potential `circuit breaker' against cavitation},\ }\href@noop {} {\bibfield  {journal} {\bibinfo  {journal} {Plant Physiol.}\ }\textbf {\bibinfo {volume} {172}},\ \bibinfo {pages} {2261} (\bibinfo {year} {2016})}\BibitemShut {NoStop}%
\bibitem [{\citenamefont {Chin}\ \emph {et~al.}(2022)\citenamefont {Chin}, \citenamefont {Guzmán‐Delgado}, \citenamefont {Sillett}, \citenamefont {Kerhoulas}, \citenamefont {R}, \citenamefont {{McElrone}},\ and\ \citenamefont {Zwieniecki}}]{chin_tracheid_2022}%
  \BibitemOpen
  \bibfield  {author} {\bibinfo {author} {\bibfnamefont {A.~R.}\ \bibnamefont {Chin}}, \bibinfo {author} {\bibfnamefont {P.}~\bibnamefont {Guzmán‐Delgado}}, \bibinfo {author} {\bibfnamefont {S.~C.}\ \bibnamefont {Sillett}}, \bibinfo {author} {\bibfnamefont {L.~P.}\ \bibnamefont {Kerhoulas}}, \bibinfo {author} {\bibfnamefont {A.~R. A.~A.}\ \bibnamefont {R}}, \bibinfo {author} {\bibnamefont {{McElrone}}},\ and\ \bibinfo {author} {\bibfnamefont {M.~A.}\ \bibnamefont {Zwieniecki}},\ }\bibfield  {title} {\bibinfo {title} {Tracheid buckling buys time, foliar water uptake pays it back: {Coordination} of leaf structure and function in tall redwood trees},\ }\href@noop {} {\bibfield  {journal} {\bibinfo  {journal} {Plant Cell Environ.}\ }\textbf {\bibinfo {volume} {45}},\ \bibinfo {pages} {2607} (\bibinfo {year} {2022})}\BibitemShut {NoStop}%
\bibitem [{\citenamefont {Wheeler}\ and\ \citenamefont {Stroock}(2008)}]{wheeler_transpiration_2008}%
  \BibitemOpen
  \bibfield  {author} {\bibinfo {author} {\bibfnamefont {T.~D.}\ \bibnamefont {Wheeler}}\ and\ \bibinfo {author} {\bibfnamefont {A.~D.}\ \bibnamefont {Stroock}},\ }\bibfield  {title} {\bibinfo {title} {The transpiration of water at negative pressures in a synthetic tree},\ }\href {https://doi.org/10.1038/nature07226} {\bibfield  {journal} {\bibinfo  {journal} {Nature}\ }\textbf {\bibinfo {volume} {455}},\ \bibinfo {pages} {208} (\bibinfo {year} {2008})}\BibitemShut {NoStop}%
\bibitem [{\citenamefont {Noblin}\ \emph {et~al.}(2008)\citenamefont {Noblin}, \citenamefont {Mahadevan}, \citenamefont {Coomaraswamy}, \citenamefont {Weitz}, \citenamefont {Holbrook},\ and\ \citenamefont {Zwieniecki}}]{noblin_optimal_2008}%
  \BibitemOpen
  \bibfield  {author} {\bibinfo {author} {\bibfnamefont {X.}~\bibnamefont {Noblin}}, \bibinfo {author} {\bibfnamefont {L.}~\bibnamefont {Mahadevan}}, \bibinfo {author} {\bibfnamefont {I.~A.}\ \bibnamefont {Coomaraswamy}}, \bibinfo {author} {\bibfnamefont {D.~A.}\ \bibnamefont {Weitz}}, \bibinfo {author} {\bibfnamefont {N.~M.}\ \bibnamefont {Holbrook}},\ and\ \bibinfo {author} {\bibfnamefont {M.~A.}\ \bibnamefont {Zwieniecki}},\ }\bibfield  {title} {\bibinfo {title} {Optimal vein density in artificial and real leaves},\ }\href@noop {} {\bibfield  {journal} {\bibinfo  {journal} {Proc. Natl. Acad. Sci.}\ }\textbf {\bibinfo {volume} {105}},\ \bibinfo {pages} {9140} (\bibinfo {year} {2008})}\BibitemShut {NoStop}%
\bibitem [{\citenamefont {Duan}\ \emph {et~al.}(2012)\citenamefont {Duan}, \citenamefont {Karnik}, \citenamefont {Lu},\ and\ \citenamefont {Majumdar}}]{duan_evaporation-induced_2012}%
  \BibitemOpen
  \bibfield  {author} {\bibinfo {author} {\bibfnamefont {C.}~\bibnamefont {Duan}}, \bibinfo {author} {\bibfnamefont {R.}~\bibnamefont {Karnik}}, \bibinfo {author} {\bibfnamefont {M.-C.}\ \bibnamefont {Lu}},\ and\ \bibinfo {author} {\bibfnamefont {A.}~\bibnamefont {Majumdar}},\ }\bibfield  {title} {\bibinfo {title} {Evaporation-induced cavitation in nanofluidic channels},\ }\href@noop {} {\bibfield  {journal} {\bibinfo  {journal} {Proc. Natl. Acad. Sci.}\ }\textbf {\bibinfo {volume} {109}},\ \bibinfo {pages} {3688} (\bibinfo {year} {2012})}\BibitemShut {NoStop}%
\bibitem [{\citenamefont {Vincent}\ \emph {et~al.}(2012)\citenamefont {Vincent}, \citenamefont {Marmottant}, \citenamefont {Quinto-Su},\ and\ \citenamefont {Ohl}}]{vincent_birth_2012}%
  \BibitemOpen
  \bibfield  {author} {\bibinfo {author} {\bibfnamefont {O.}~\bibnamefont {Vincent}}, \bibinfo {author} {\bibfnamefont {P.}~\bibnamefont {Marmottant}}, \bibinfo {author} {\bibfnamefont {P.~A.}\ \bibnamefont {Quinto-Su}},\ and\ \bibinfo {author} {\bibfnamefont {C.-D.}\ \bibnamefont {Ohl}},\ }\bibfield  {title} {\bibinfo {title} {Birth and growth of cavitation bubbles within water under tension confined in a simple synthetic tree},\ }\href@noop {} {\bibfield  {journal} {\bibinfo  {journal} {Phys. Rev. Lett.}\ }\textbf {\bibinfo {volume} {108}},\ \bibinfo {pages} {184502} (\bibinfo {year} {2012})}\BibitemShut {NoStop}%
\bibitem [{\citenamefont {Vincent}\ \emph {et~al.}(2014)\citenamefont {Vincent}, \citenamefont {Sessoms}, \citenamefont {Huber}, \citenamefont {Guioth},\ and\ \citenamefont {Stroock}}]{vincent_drying_2014}%
  \BibitemOpen
  \bibfield  {author} {\bibinfo {author} {\bibfnamefont {O.}~\bibnamefont {Vincent}}, \bibinfo {author} {\bibfnamefont {D.~A.}\ \bibnamefont {Sessoms}}, \bibinfo {author} {\bibfnamefont {E.~J.}\ \bibnamefont {Huber}}, \bibinfo {author} {\bibfnamefont {J.}~\bibnamefont {Guioth}},\ and\ \bibinfo {author} {\bibfnamefont {A.~D.}\ \bibnamefont {Stroock}},\ }\bibfield  {title} {\bibinfo {title} {Drying by {Cavitation} and {Poroelastic} {Relaxations} in {Porous} {Media} with {Macroscopic} {Pores} {Connected} by {Nanoscale} {Throats}},\ }\href {https://doi.org/10.1103/PhysRevLett.113.134501} {\bibfield  {journal} {\bibinfo  {journal} {Phys. Rev. Lett.}\ }\textbf {\bibinfo {volume} {113}},\ \bibinfo {pages} {134501} (\bibinfo {year} {2014})}\BibitemShut {NoStop}%
\bibitem [{\citenamefont {Bruning}\ \emph {et~al.}(2019{\natexlab{a}})\citenamefont {Bruning}, \citenamefont {Costalonga}, \citenamefont {Snoeijer},\ and\ \citenamefont {Marín}}]{bruning_turning_2019}%
  \BibitemOpen
  \bibfield  {author} {\bibinfo {author} {\bibfnamefont {M.~A.}\ \bibnamefont {Bruning}}, \bibinfo {author} {\bibfnamefont {M.}~\bibnamefont {Costalonga}}, \bibinfo {author} {\bibfnamefont {J.~H.}\ \bibnamefont {Snoeijer}},\ and\ \bibinfo {author} {\bibfnamefont {A.}~\bibnamefont {Marín}},\ }\bibfield  {title} {\bibinfo {title} {Turning drops into bubbles: cavitation by vapor diffusion through elastic networks},\ }\href@noop {} {\bibfield  {journal} {\bibinfo  {journal} {Phys. Rev. Lett.}\ }\textbf {\bibinfo {volume} {123}},\ \bibinfo {pages} {214501} (\bibinfo {year} {2019}{\natexlab{a}})}\BibitemShut {NoStop}%
\bibitem [{\citenamefont {Dollet}\ \emph {et~al.}(2019)\citenamefont {Dollet}, \citenamefont {Louf}, \citenamefont {Alonzo}, \citenamefont {Jensen},\ and\ \citenamefont {Marmottant}}]{dollet_drying_2019}%
  \BibitemOpen
  \bibfield  {author} {\bibinfo {author} {\bibfnamefont {B.}~\bibnamefont {Dollet}}, \bibinfo {author} {\bibfnamefont {J.~F.}\ \bibnamefont {Louf}}, \bibinfo {author} {\bibfnamefont {M.}~\bibnamefont {Alonzo}}, \bibinfo {author} {\bibfnamefont {K.~H.}\ \bibnamefont {Jensen}},\ and\ \bibinfo {author} {\bibfnamefont {P.}~\bibnamefont {Marmottant}},\ }\bibfield  {title} {\bibinfo {title} {Drying of channels by evaporation through a permeable medium},\ }\href@noop {} {\bibfield  {journal} {\bibinfo  {journal} {J. R. Soc. Interface}\ }\textbf {\bibinfo {volume} {16}},\ \bibinfo {pages} {20180690} (\bibinfo {year} {2019})}\BibitemShut {NoStop}%
\bibitem [{\citenamefont {Dollet}\ \emph {et~al.}(2021)\citenamefont {Dollet}, \citenamefont {Encarnación}, \citenamefont {Gautier},\ and\ \citenamefont {Marmottant}}]{dollet_drying_2021}%
  \BibitemOpen
  \bibfield  {author} {\bibinfo {author} {\bibfnamefont {B.}~\bibnamefont {Dollet}}, \bibinfo {author} {\bibfnamefont {K.~N.~C.}\ \bibnamefont {Encarnación}}, \bibinfo {author} {\bibfnamefont {R.}~\bibnamefont {Gautier}},\ and\ \bibinfo {author} {\bibfnamefont {P.}~\bibnamefont {Marmottant}},\ }\bibfield  {title} {\bibinfo {title} {Drying by pervaporation in elementary channel networks},\ }\href@noop {} {\bibfield  {journal} {\bibinfo  {journal} {J. Fluid Mech.}\ }\textbf {\bibinfo {volume} {906}},\ \bibinfo {pages} {A6} (\bibinfo {year} {2021})}\BibitemShut {NoStop}%
\bibitem [{\citenamefont {Vincent}\ \emph {et~al.}(2024)\citenamefont {Vincent}, \citenamefont {Tassin}, \citenamefont {Huber},\ and\ \citenamefont {Stroock}}]{vincent_tunable_2024}%
  \BibitemOpen
  \bibfield  {author} {\bibinfo {author} {\bibfnamefont {O.}~\bibnamefont {Vincent}}, \bibinfo {author} {\bibfnamefont {T.}~\bibnamefont {Tassin}}, \bibinfo {author} {\bibfnamefont {E.~J.}\ \bibnamefont {Huber}},\ and\ \bibinfo {author} {\bibfnamefont {A.~D.}\ \bibnamefont {Stroock}},\ }\bibfield  {title} {\bibinfo {title} {Tunable transport in bidisperse porous materials with vascular structure},\ }\href {https://doi.org/10.1103/PhysRevFluids.9.064202} {\bibfield  {journal} {\bibinfo  {journal} {Phys. Rev. Fluids}\ }\textbf {\bibinfo {volume} {9}},\ \bibinfo {pages} {064202} (\bibinfo {year} {2024})}\BibitemShut {NoStop}%
\bibitem [{\citenamefont {Keiser}\ \emph {et~al.}(2022)\citenamefont {Keiser}, \citenamefont {Marmottant},\ and\ \citenamefont {Dollet}}]{keiser_intermittent_2022}%
  \BibitemOpen
  \bibfield  {author} {\bibinfo {author} {\bibfnamefont {L.}~\bibnamefont {Keiser}}, \bibinfo {author} {\bibfnamefont {P.}~\bibnamefont {Marmottant}},\ and\ \bibinfo {author} {\bibfnamefont {B.}~\bibnamefont {Dollet}},\ }\bibfield  {title} {\bibinfo {title} {Intermittent air invasion in pervaporating compliant microchannels},\ }\href {https://doi.org/https://doi.org/10.1017/jfm.2022.733} {\bibfield  {journal} {\bibinfo  {journal} {J. Fluid Mech.}\ }\textbf {\bibinfo {volume} {948}},\ \bibinfo {pages} {A52} (\bibinfo {year} {2022})}\BibitemShut {NoStop}%
\bibitem [{\citenamefont {Keiser}\ \emph {et~al.}(2024)\citenamefont {Keiser}, \citenamefont {Dollet},\ and\ \citenamefont {Marmottant}}]{keiser_embolism_2024}%
  \BibitemOpen
  \bibfield  {author} {\bibinfo {author} {\bibfnamefont {L.}~\bibnamefont {Keiser}}, \bibinfo {author} {\bibfnamefont {B.}~\bibnamefont {Dollet}},\ and\ \bibinfo {author} {\bibfnamefont {P.}~\bibnamefont {Marmottant}},\ }\bibfield  {title} {\bibinfo {title} {Embolism propagation in \textit{{Adiantum}} leaves and in a biomimetic system with constrictions},\ }\href {https://doi.org/10.1098/rsif.2024.0103} {\bibfield  {journal} {\bibinfo  {journal} {Journal of The Royal Society Interface}\ }\textbf {\bibinfo {volume} {21}},\ \bibinfo {pages} {20240103} (\bibinfo {year} {2024})}\BibitemShut {NoStop}%
\bibitem [{\citenamefont {Bruus}(2008)}]{bruus_theoretical_2008}%
  \BibitemOpen
  \bibfield  {author} {\bibinfo {author} {\bibfnamefont {H.}~\bibnamefont {Bruus}},\ }\href@noop {} {\emph {\bibinfo {title} {Theoretical {Microfluidics}}}}\ (\bibinfo  {publisher} {Oxford University Press},\ \bibinfo {year} {2008})\BibitemShut {NoStop}%
\bibitem [{\citenamefont {Wong}\ \emph {et~al.}(1992)\citenamefont {Wong}, \citenamefont {Morris},\ and\ \citenamefont {Radke}}]{wong1992three}%
  \BibitemOpen
  \bibfield  {author} {\bibinfo {author} {\bibfnamefont {H.}~\bibnamefont {Wong}}, \bibinfo {author} {\bibfnamefont {S.}~\bibnamefont {Morris}},\ and\ \bibinfo {author} {\bibfnamefont {C.}~\bibnamefont {Radke}},\ }\bibfield  {title} {\bibinfo {title} {Three-dimensional menisci in polygonal capillaries},\ }\href@noop {} {\bibfield  {journal} {\bibinfo  {journal} {Journal of Colloid and Interface Science}\ }\textbf {\bibinfo {volume} {148}},\ \bibinfo {pages} {317} (\bibinfo {year} {1992})}\BibitemShut {NoStop}%
\bibitem [{Not()}]{NoteOnPc}%
  \BibitemOpen
  \href@noop {} {}\bibinfo {note} {The latter condition ensures the continuity, between the discrete and the continuum model, of the volume injection in the remaining wet channel series due to the movement of the interface. By definition of the continuum approximation, noting $\langle . \rangle$ the average for many channel draining observations, the water volume flux coming from the embolization of the channels is: $\left\langle S_\mathrm{c}l/\Delta t_\mathrm{tot} \right\rangle $ with $S_\mathrm{c} = S_0 + Cp_\mathrm{c}/l$. Given that $\left\langle l/\Delta t_\mathrm{tot} \right\rangle = \dot{L}$, we indeed find that we have to impose the boundary condition $p = p_\mathrm{c}$ at the interface to conserve the volume flux between the discrete and continuous model.}\BibitemShut {Stop}%
\bibitem [{\citenamefont {Encarnación}\ \emph {et~al.}(2021)\citenamefont {Encarnación}, \citenamefont {Marmottant},\ and\ \citenamefont {Dollet}}]{encarnacion_pervaporation-induced_2021}%
  \BibitemOpen
  \bibfield  {author} {\bibinfo {author} {\bibfnamefont {K.~N.~C.}\ \bibnamefont {Encarnación}}, \bibinfo {author} {\bibfnamefont {P.}~\bibnamefont {Marmottant}},\ and\ \bibinfo {author} {\bibfnamefont {B.}~\bibnamefont {Dollet}},\ }\bibfield  {title} {\bibinfo {title} {Pervaporation-induced drying in networks of channels of variable width},\ }\href@noop {} {\bibfield  {journal} {\bibinfo  {journal} {Microfluid. Nanofluid.}\ }\textbf {\bibinfo {volume} {25}},\ \bibinfo {pages} {71} (\bibinfo {year} {2021})}\BibitemShut {NoStop}%
\bibitem [{con()}]{continuous}%
  \BibitemOpen
  \href@noop {} {}\bibinfo {note} {Note that the continuous model was solved using a simplified version of the pervaporation rate, only keeping the linear dependence of $j$ with the $w$ and neglecting the corrective terms of Eq. \ref{eq:pervap}.}\BibitemShut {Stop}%
\bibitem [{\citenamefont {Derr}\ \emph {et~al.}(2020)\citenamefont {Derr}, \citenamefont {Fronk}, \citenamefont {Weber}, \citenamefont {Mahadevan}, \citenamefont {Rycroft},\ and\ \citenamefont {Mahadevan}}]{derr_flow-driven_2020}%
  \BibitemOpen
  \bibfield  {author} {\bibinfo {author} {\bibfnamefont {N.~J.}\ \bibnamefont {Derr}}, \bibinfo {author} {\bibfnamefont {D.~C.}\ \bibnamefont {Fronk}}, \bibinfo {author} {\bibfnamefont {C.~A.}\ \bibnamefont {Weber}}, \bibinfo {author} {\bibfnamefont {A.}~\bibnamefont {Mahadevan}}, \bibinfo {author} {\bibfnamefont {C.~H.}\ \bibnamefont {Rycroft}},\ and\ \bibinfo {author} {\bibfnamefont {L.}~\bibnamefont {Mahadevan}},\ }\bibfield  {title} {\bibinfo {title} {Flow-{Driven} {Branching} in a {Frangible} {Porous} {Medium}},\ }\href {https://doi.org/10.1103/PhysRevLett.125.158002} {\bibfield  {journal} {\bibinfo  {journal} {Phys. Rev. Lett.}\ }\textbf {\bibinfo {volume} {125}},\ \bibinfo {pages} {158002} (\bibinfo {year} {2020})}\BibitemShut {NoStop}%
\bibitem [{\citenamefont {Zareei}\ \emph {et~al.}(2022)\citenamefont {Zareei}, \citenamefont {Pan},\ and\ \citenamefont {Amir}}]{Zareei2022}%
  \BibitemOpen
  \bibfield  {author} {\bibinfo {author} {\bibfnamefont {A.}~\bibnamefont {Zareei}}, \bibinfo {author} {\bibfnamefont {D.}~\bibnamefont {Pan}},\ and\ \bibinfo {author} {\bibfnamefont {A.}~\bibnamefont {Amir}},\ }\bibfield  {title} {\bibinfo {title} {Temporal evolution of erosion in pore networks: From homogenization to instability},\ }\href {https://doi.org/10.1103/PhysRevLett.128.234501} {\bibfield  {journal} {\bibinfo  {journal} {Phys. Rev. Lett.}\ }\textbf {\bibinfo {volume} {128}},\ \bibinfo {pages} {234501} (\bibinfo {year} {2022})}\BibitemShut {NoStop}%
\bibitem [{\citenamefont {Fancher}\ and\ \citenamefont {Katifori}(2022)}]{fancher_mechanical_2022}%
  \BibitemOpen
  \bibfield  {author} {\bibinfo {author} {\bibfnamefont {S.}~\bibnamefont {Fancher}}\ and\ \bibinfo {author} {\bibfnamefont {E.}~\bibnamefont {Katifori}},\ }\bibfield  {title} {\bibinfo {title} {Mechanical response in elastic fluid flow networks},\ }\href {https://doi.org/10.1103/PhysRevFluids.7.013101} {\bibfield  {journal} {\bibinfo  {journal} {Phys. Rev. Fluids}\ }\textbf {\bibinfo {volume} {7}},\ \bibinfo {pages} {013101} (\bibinfo {year} {2022})}\BibitemShut {NoStop}%
\bibitem [{\citenamefont {Winn}\ and\ \citenamefont {Katifori}(2024)}]{winn_operating_2024}%
  \BibitemOpen
  \bibfield  {author} {\bibinfo {author} {\bibfnamefont {A.}~\bibnamefont {Winn}}\ and\ \bibinfo {author} {\bibfnamefont {E.}~\bibnamefont {Katifori}},\ }\bibfield  {title} {\bibinfo {title} {Operating principles of peristaltic pumping through a dense array of valves},\ }\href {https://doi.org/10.1017/jfm.2024.480} {\bibfield  {journal} {\bibinfo  {journal} {Journal of Fluid Mechanics}\ }\textbf {\bibinfo {volume} {989}},\ \bibinfo {pages} {A18} (\bibinfo {year} {2024})}\BibitemShut {NoStop}%
\bibitem [{\citenamefont {Bacchin}\ \emph {et~al.}(2022)\citenamefont {Bacchin}, \citenamefont {Leng},\ and\ \citenamefont {Salmon}}]{bacchin_microfluidic_2022}%
  \BibitemOpen
  \bibfield  {author} {\bibinfo {author} {\bibfnamefont {P.}~\bibnamefont {Bacchin}}, \bibinfo {author} {\bibfnamefont {J.}~\bibnamefont {Leng}},\ and\ \bibinfo {author} {\bibfnamefont {J.~B.}\ \bibnamefont {Salmon}},\ }\bibfield  {title} {\bibinfo {title} {Microfluidic evaporation, pervaporation, and osmosis: {From} passive pumping to solute concentration},\ }\href@noop {} {\bibfield  {journal} {\bibinfo  {journal} {Chem. Rev.}\ }\textbf {\bibinfo {volume} {122}},\ \bibinfo {pages} {6938} (\bibinfo {year} {2022})}\BibitemShut {NoStop}%
\bibitem [{\citenamefont {Gauci}\ \emph {et~al.}(2025)\citenamefont {Gauci}, \citenamefont {Jami}, \citenamefont {Keiser}, \citenamefont {Cohen},\ and\ \citenamefont {Noblin}}]{gauci2025channel}%
  \BibitemOpen
  \bibfield  {author} {\bibinfo {author} {\bibfnamefont {F.-X.}\ \bibnamefont {Gauci}}, \bibinfo {author} {\bibfnamefont {L.}\ \bibnamefont {Jami}}, \bibinfo {author} {\bibfnamefont {L.}\ \bibnamefont {Keiser}}, \bibinfo {author} {\bibfnamefont {C.}\ \bibnamefont {Cohen}},\ and\ \bibinfo {author} {\bibfnamefont {X.}\ \bibnamefont {Noblin}},}%
  \bibfield  {title} {\bibinfo {title} { Channel deformations during elastocapillary spreading of gaseous embolisms in biomimetic leaves},}%
  \href {https://doi.org/10.1098/rsfs.2024.0060} {\bibfield  {journal} {\bibinfo  {journal} { Interface Focus}\ }\textbf {\bibinfo {volume} {15}},\ \bibinfo {pages} {20240060} (\bibinfo {year} {2025})}
  \BibitemShut {NoStop}%
\bibitem [{\citenamefont {Bruning}\ \emph {et~al.}(2019{\natexlab{b}})\citenamefont {Bruning}, \citenamefont {Costalonga}, \citenamefont {Snoeijer},\ and\ \citenamefont {Marin}}]{bruning_turning_2019-1}%
  \BibitemOpen
  \bibfield  {author} {\bibinfo {author} {\bibfnamefont {M.}~\bibnamefont {Bruning}}, \bibinfo {author} {\bibfnamefont {M.}~\bibnamefont {Costalonga}}, \bibinfo {author} {\bibfnamefont {J.}~\bibnamefont {Snoeijer}},\ and\ \bibinfo {author} {\bibfnamefont {A.}~\bibnamefont {Marin}},\ }\bibfield  {title} {\bibinfo {title} {Turning {Drops} into {Bubbles}: {Cavitation} by {Vapor} {Diffusion} through {Elastic} {Networks}},\ }\href {https://doi.org/10.1103/PhysRevLett.123.214501} {\bibfield  {journal} {\bibinfo  {journal} {Phys. Rev. Lett.}\ }\textbf {\bibinfo {volume} {123}},\ \bibinfo {pages} {214501} (\bibinfo {year} {2019}{\natexlab{b}})}\BibitemShut {NoStop}%
\bibitem [{\citenamefont {Pingulkar}\ \emph {et~al.}(2024)\citenamefont {Pingulkar}, \citenamefont {Ayela},\ and\ \citenamefont {Salmon}}]{pingulkar_pervaporation-driven_2024}%
  \BibitemOpen
  \bibfield  {author} {\bibinfo {author} {\bibfnamefont {H.}~\bibnamefont {Pingulkar}}, \bibinfo {author} {\bibfnamefont {C.}~\bibnamefont {Ayela}},\ and\ \bibinfo {author} {\bibfnamefont {J.-B.}\ \bibnamefont {Salmon}},\ }\bibfield  {title} {\bibinfo {title} {Pervaporation-driven electrokinetic energy harvesting using poly(dimethylsiloxane) microfluidic chips},\ }\href {https://doi.org/10.1039/D4LC00831F} {\bibfield  {journal} {\bibinfo  {journal} {Lab on a Chip}\ }\textbf {\bibinfo {volume} {24}},\ \bibinfo {pages} {5328} (\bibinfo {year} {2024})}\BibitemShut {NoStop}%
\bibitem [{\citenamefont {Blackman}\ and\ \citenamefont {Brodribb}(2011)}]{blackman_two_2011}%
  \BibitemOpen
  \bibfield  {author} {\bibinfo {author} {\bibfnamefont {C.~J.}\ \bibnamefont {Blackman}}\ and\ \bibinfo {author} {\bibfnamefont {T.~J.}\ \bibnamefont {Brodribb}},\ }\bibfield  {title} {\bibinfo {title} {Two measures of leaf capacitance: insights into the water transport pathway and hydraulic conductance in leaves},\ }\href@noop {} {\bibfield  {journal} {\bibinfo  {journal} {Functional Plant Biology}\ }\textbf {\bibinfo {volume} {38}},\ \bibinfo {pages} {118} (\bibinfo {year} {2011})}\BibitemShut {NoStop}%
\bibitem [{\citenamefont {Hölttä}\ \emph {et~al.}(2009)\citenamefont {Hölttä}, \citenamefont {Cochard}, \citenamefont {Nikinmaa},\ and\ \citenamefont {Mencuccini}}]{holtta_capacitive_2009}%
  \BibitemOpen
  \bibfield  {author} {\bibinfo {author} {\bibfnamefont {T.}~\bibnamefont {Hölttä}}, \bibinfo {author} {\bibfnamefont {H.}~\bibnamefont {Cochard}}, \bibinfo {author} {\bibfnamefont {E.}~\bibnamefont {Nikinmaa}},\ and\ \bibinfo {author} {\bibfnamefont {M.}~\bibnamefont {Mencuccini}},\ }\bibfield  {title} {\bibinfo {title} {Capacitive effect of cavitation in xylem conduits: results from a dynamic model},\ }\href {https://doi.org/10.1111/j.1365-3040.2008.01894.x} {\bibfield  {journal} {\bibinfo  {journal} {Plant, Cell \& Environment}\ }\textbf {\bibinfo {volume} {32}},\ \bibinfo {pages} {10} (\bibinfo {year} {2009})}\BibitemShut {NoStop}%
\bibitem [{\citenamefont {Meinzer}\ \emph {et~al.}(2004)\citenamefont {Meinzer}, \citenamefont {James},\ and\ \citenamefont {Goldstein}}]{meinzer2004dynamics}%
\BibitemOpen
\bibfield  {author} {\bibinfo {author} {\bibfnamefont {F.~C.}\ \bibnamefont {Meinzer}}, \bibinfo {author} {\bibfnamefont {S.~A.}\ \bibnamefont {James}},\ and\ \bibinfo {author} {\bibfnamefont {G.}\ \bibnamefont {Goldstein}},\ }\bibfield  {title} {\bibinfo {title} {Dynamics of transpiration, sap flow and use of stored water in tropical forest canopy trees},\ }\bibfield  {journal} {\bibinfo  {journal} {Tree Physiology}\ }\textbf {\bibinfo {volume} {24}},\ \bibinfo {pages} {901--909} (\bibinfo {year}{2004})\BibitemShut {NoStop}\bibitem [{\citenamefont {Phillips}\ \emph {et~al.}(1997)\citenamefont {Phillips}, \citenamefont {Nagchaudhuri}, \citenamefont {Oren},\ and\ \citenamefont {Katul}}]{phillips1997time}%
\BibitemOpen
\bibfield  {author} {\bibinfo {author} {\bibfnamefont {N.}\ \bibnamefont {Phillips}}, \bibinfo {author} {\bibfnamefont {A.}\ \bibnamefont {Nagchaudhuri}}, \bibinfo {author} {\bibfnamefont {R.}\ \bibnamefont {Oren}},\ and\ \bibinfo {author} {\bibfnamefont {G.}\ \bibnamefont {Katul}},\ }\bibfield  {title} {\bibinfo {title} {Time constant for water transport in loblolly pine trees estimated from time series of evaporative demand and stem sapflow},\ }\bibfield  {journal} {\bibinfo  {journal} {Trees}\ }\textbf {\bibinfo {volume} {11}},\ \bibinfo {pages} {412--419} (\bibinfo {year} {1997})\BibitemShut{NoStop}%
\bibitem [{\citenamefont {Luo}\ \emph {et~al.}(2021)\citenamefont {Luo}, \citenamefont {Ho}, \citenamefont {Helliker},\ and\ \citenamefont {Katifori}}]{luo2021leaf}%
\BibitemOpen
\bibfield  {author} {\bibinfo {author} {\bibfnamefont {Y.}~\bibnamefont {Luo}}, \bibinfo {author} {\bibfnamefont {C.-L.}\ \bibnamefont {Ho}}, \bibinfo {author} {\bibfnamefont {B.~R.}\ \bibnamefont {Helliker}},\ and\ \bibinfo {author} {\bibfnamefont {E.}\ \bibnamefont {Katifori}},\ }\bibfield  {title} {\bibinfo {title} {Leaf water storage and robustness to intermittent drought: A spatially explicit capacitive model for leaf hydraulics},\ }\bibfield  {journal} {\bibinfo  {journal} {Frontiers in Plant Science}\ }\textbf {\bibinfo {volume} {12}},\ \bibinfo {pages} {725995} (\bibinfo {year} {2021})\BibitemShut{NoStop}\bibitem [{\citenamefont {Zhuang}\ \emph {et~al.}(2014)\citenamefont {Zhuang}, \citenamefont {Yu},\ and\ \citenamefont {Nakayama}}]{zhuang2014series}%
\BibitemOpen
\bibfield  {author} {\bibinfo {author} {\bibfnamefont {J.}\ \bibnamefont {Zhuang}}, \bibinfo {author} {\bibfnamefont {G.-R.}\ \bibnamefont {Yu}},\ and\ \bibinfo {author} {\bibfnamefont {K.}\ \bibnamefont {Nakayama}},\ }\bibfield  {title} {\bibinfo {title} {A series {RCL} circuit theory for analyzing non-steady-state water uptake of maize plants},\ }\bibfield  {journal} {\bibinfo  {journal} {Scientific Reports}\ }\textbf {\bibinfo {volume} {4}},\ \bibinfo {pages} {6720} (\bibinfo {year} {2014})\BibitemShut {NoStop}%
\bibitem [{\citenamefont {Zhang}\ \emph {et~al.}(2024)\citenamefont {Zhang}, \citenamefont {Pereira}, \citenamefont {Kaack}, \citenamefont {Liu},\ and\ \citenamefont {Jansen}}]{zhang_gold_2024}%
  \BibitemOpen
  \bibfield  {author} {\bibinfo {author} {\bibfnamefont {Y.}~\bibnamefont {Zhang}}, \bibinfo {author} {\bibfnamefont {L.}~\bibnamefont {Pereira}}, \bibinfo {author} {\bibfnamefont {L.}~\bibnamefont {Kaack}}, \bibinfo {author} {\bibfnamefont {J.}~\bibnamefont {Liu}},\ and\ \bibinfo {author} {\bibfnamefont {S.}~\bibnamefont {Jansen}},\ }\bibfield  {title} {\bibinfo {title} {Gold perfusion experiments support the multi-layered, mesoporous nature of intervessel pit membranes in angiosperm xylem},\ }\href@noop {} {\bibfield  {journal} {\bibinfo  {journal} {New Phytologist}\ } (\bibinfo {year} {2024})}\BibitemShut {NoStop}%
\bibitem [{\citenamefont {Pittermann}\ \emph {et~al.}(2005)\citenamefont {Pittermann}, \citenamefont {Sperry}, \citenamefont {Hacke}, \citenamefont {Wheeler},\ and\ \citenamefont {Sikkema}}]{pittermann_torus-margo_2005}%
  \BibitemOpen
  \bibfield  {author} {\bibinfo {author} {\bibfnamefont {J.}~\bibnamefont {Pittermann}}, \bibinfo {author} {\bibfnamefont {J.~S.}\ \bibnamefont {Sperry}}, \bibinfo {author} {\bibfnamefont {U.~G.}\ \bibnamefont {Hacke}}, \bibinfo {author} {\bibfnamefont {J.~K.}\ \bibnamefont {Wheeler}},\ and\ \bibinfo {author} {\bibfnamefont {E.~H.}\ \bibnamefont {Sikkema}},\ }\bibfield  {title} {\bibinfo {title} {Torus-{Margo} {Pits} {Help} {Conifers} {Compete} with {Angiosperms}},\ }\href {https://doi.org/10.1126/science.1120479} {\bibfield  {journal} {\bibinfo  {journal} {Science}\ }\textbf {\bibinfo {volume} {310}},\ \bibinfo {pages} {1924} (\bibinfo {year} {2005})}\BibitemShut {NoStop}%
  \bibitem [{\citenamefont {Jami}\ \emph {et~al.}(2025)\citenamefont {Jami}, \citenamefont {Gauci}, \citenamefont {Cohen}, \citenamefont {Noblin},\ and\ \citenamefont {Keiser}}]{zenodo17485963}%
  \BibitemOpen
  \bibfield  {author} {\bibinfo {author} {\bibfnamefont {L.}\ \bibnamefont {Jami}}, \bibinfo {author} {\bibfnamefont {F.-X.}\ \bibnamefont {Gauci}}, \bibinfo {author} {\bibfnamefont {C.}\ \bibnamefont {Cohen}}, \bibinfo {author} {\bibfnamefont {X.}\ \bibnamefont {Noblin}},\ and\ \bibinfo {author} {\bibfnamefont {L.}\ \bibnamefont {Keiser}},}%
  \bibfield  {title} {\bibinfo {title} { Supplementary material for ``Nonlinear dynamics of air invasion in one-dimensional compliant fluid networks'' [Data set]},}%
  \href {https://doi.org/10.5281/zenodo.17485963} {\bibfield  {journal} {\bibinfo  {journal}  { Zenodo}\ }} (\bibinfo {year} {2025}),\ \bibinfo {note} {in Physical Review E (doi:10.5281/zenodo.17485963)}%
  \BibitemShut {NoStop}%
\end{thebibliography}
\end{document}